%% file: Qexp-ABSDEJ-NV-WK.tex
\documentclass[a4,11pt]{article}
\usepackage{graphicx} 
\usepackage{float}
\usepackage{amsmath,amsthm}
\usepackage{amsfonts}
\usepackage{latexsym}
\usepackage{amssymb}
\usepackage{setspace}
\usepackage{comment}

\makeatletter
 
  \@addtoreset{equation}{section}
 \makeatother

\begin{document}
\title{Anticipated Backward SDEs with Jumps and \\ quadratic-exponential growth drivers
~\footnote{
 Accepted for publication in Stochastics and Dynamics.
All the contents expressed in this research are solely those of the author and do not represent any views or 
opinions of any institutions. 
}}
%\\ \Large{\it{--Implications of asymmetric CSA and suboptimal strategies--}}

\author{Masaaki Fujii\footnote{Quantitative Finance Course, Graduate School of Economics, The University of Tokyo.
mfujii@e.u-tokyo.ac.jp}
~~\&~~Akihiko Takahashi\footnote{Quantitative Finance Course, Graduate School of Economics, The University of Tokyo. 
akihikot@e.u-tokyo.ac.jp} 
}
\date{ 
%First version May 6, 2017 \\
This version: July 9, 2018}
\maketitle

%%%%%%    TEXT START    %%%%%%

%%%%%%      Macros      %%%%%%
\input{Fmacro-2015.tex}

%\newtheorem{definition}{Definition}[section]
%\newtheorem{assumption}{$[$ A}
%\newtheorem{condition}{$[$ C}
%\newtheorem{lemma}{Lemma}[section]
%\newtheorem{proposition}{Proposition}[section]
%\newtheorem{theorem}{Theorem}[section]
%\newtheorem{remark}{Remark}
%\newtheorem{example}{Example}
%\newtheorem{corollary}{Corollary}[section]

\def\E{{\bf E}}

\def\cadlag{{c\`adl\`ag~}}

\def\calf{{\cal F}}
\def\wt{\widetilde}
\def\mbb{\mathbb}
\def\ol{\overline}
\def\ul{\underline}
\def\sign{{\rm{sign}}}

\def\tp{{t^\prime}}
\def\sp{{s^\prime}}
\def\xp{{x^\prime}}
\def\bq{{\bar{q}}}
\def\p{\prime}
\def\ep{\epsilon}
\def\del{\delta}
\def\Del{\Delta}
\def\part{\partial}

\def\nn{\nonumber}
\def\be{\begin{equation}}
\def\ee{\end{equation}}
\def\bea{\begin{eqnarray}}
\def\eea{\end{eqnarray}}
\def\beas{\begin{eqnarray*}}
\def\eeas{\end{eqnarray*}}
\def\l{\left}
\def\r{\right}

\def\bull{$\bullet~$}

\newcommand{\Slash}[1]{{\ooalign{\hfil/\hfil\crcr$#1$}}}
\vspace{2mm}

%%%%%%%%%%%%%%%%%%%%%%%%%%%%%%
\begin{abstract}
%%%%%%%%%%%%%%%%%%%%%%%%%%%%%% 
In this paper, we study a class of Anticipated Backward Stochastic Differential Equations (ABSDE) with jumps. 
The solution of the ABSDE is a triple $(Y,Z,\psi)$ where $Y$ is a semimartingale, 
and $(Z,\psi)$ are the diffusion and jump coefficients. We allow the driver of the ABSDE 
to have linear growth on the uniform norm of $Y$'s future paths,
as well as quadratic and exponential growth on the spot values of $(Z,\psi)$, respectively.
The existence of the unique solution is proved for Markovian and non-Markovian settings with 
different structural assumptions on the driver. In the former case, some regularities on $(Z,\psi)$ with respect to 
the forward process are also obtained.

%%%%%%%%%%%%%%%%%%%%%%%%%%%%%%%
\end{abstract}
\vspace{3mm}
%%%%%%%%%%%%%%%%%%%%%%%%%%%%%%%%%$

%%%%%%%%%%%%%%%%%%%%%%%%%%%%%%%%%
\section{Introduction}
%%%%%%%%%%%%%%%%%%%%%%%%%%%%%%%%%
As a powerful probabilistic tool to analyze general control problems, non-linear partial differential equations
as well as many newly appeared financial problems, backward stochastic differential equations (BSDEs)
have attracted strong research interests since the pioneering works of Bismut (1973)~\cite{Bismut}
and Pardoux \& Peng (1990)~\cite{Pardoux-Peng}.

Recently, Peng \& Yang (2009)~\cite{Peng-Yang} introduced a new class, so-called anticipated (or time-advanced) BSDEs,
where the drivers are dependent on the conditional expectations of the future paths of the solutions.
They originally appeared as adjoint processes when dealing with optimal control problems on delayed systems.
Since then various generalizations have been studied by many authors:
Oksendal et al. (2011)~\cite{Oksendal} dealt with a control problem on delayed systems with jumps,
Pamen (2015)~\cite{Pamen} a stochastic differential game with delay,
Xu (2011)~\cite{Xu}, Yang \& Elliott (2013)~\cite{Yang-Elliott} studied some generalizations
and conditions for the comparison principle to hold.
Jeanblac et al. (2016)~\cite{Jeanblanc-Lim} studied anticipated BSDEs under
a setting of progressive enlargement of filtration.
The importance of anticipated BSDEs for financial applications  is likely to grow in the coming years because of the 
set of new regulations (in particular, the margin rule on the independent amount). They require the 
financial firms to adjust the collateral (or capital) amount based on the expected future  maximum loss, exposure
or the variability of the mark-to-market, which naturally makes the drivers of the pricing BSDEs dependent on 
the expected future paths of the portfolio values.

In this paper, we are interested in anticipated BSDEs with jumps and quadratic-exponential growth drivers.
Although the properties of Lipschitz ABSDEs have been well established, the ABSDEs with
quadratic growth generators have not yet appeared in the literature.
In addition to the pure mathematical interests, 
the quadratic growth (exponential growth in the presence of jumps) BSDEs have many applications. In particular, they 
arise in the context of utility optimization
with exponential or power utility functions and the associated indifference valuation,  
or questions related to risk minimization for the entropic risk measure.
They also arise in a class of recursive utilities 
introduced by Epstein \& Zin (1989)~\cite{Epstein-Zin} where the investor penalizes the variance of the value function.
Their model and its variants have found many applications in economic theory. 
Once the investor assigns a cost or benefit to the expected value of a future path,
which looks almost inevitable due to the new financial regulations,
the resultant recursive utility, which corresponds to the driver of the associated BSDE, 
starts to involve an anticipated component.
In this work,  we deal with the anticipated BSDEs with jumps of the following form:
\bea
Y_t=\xi+\int_t^T \mbb{E}_{\calf_r}f\Bigl(r,(Y_v)_{v\in[r,T]},Y_r,Z_r,\psi_r\Bigr)dr-\int_t^T Z_r dW_r
-\int_t^T\int_E \psi_r(e)\wt{\mu}(dr,de)\nn
\eea
where the driver $f(t,\cdot)$ is allowed to have  linear growth in $\sup_{v\in[t,T]}|Y_v|$, quadratic in $Z_t$
and exponential growth in the jump coefficients $\psi_t$.
This will be the necessary first step toward the understanding of the general problems
involving non-Lipschitz generators with anticipated components and its applications to the various problems
mentioned above.

For the (non-anticipated) BSDEs with quadratic growth drivers,  the first breakthrough was made by Kobylanski (2000)~\cite{Kobylanski}
and then followed by many researchers for its generalization and applications. In the presence of jumps, in particular, they were studied by 
Becherer (2006)~\cite{Becherer}, Morlais (2010)~\cite{Morlais}, Ngoupeyou (2010)~\cite{Ngoupeyou}, Cohen \& Elliott (2015)~\cite{Cohen-Elliott},
Kazi-Tani et al. (2015)~\cite{Kazi}, Antonelli \& Mancini (2016)~\cite{Antonelli}, El Karoui et al. (2016)~\cite{ElKaroui-Matoussi}
and Fujii \& Takahashi (2017)~\cite{FT-Qexp} with varying generality.
An important common tool is the so called $A_\Gamma$-condition~\cite{Barles,Royer} necessary to make 
the comparison principle to hold in the presence of jumps, which is then used to create
a monotone sequence of regularized BSDEs.

Although $A_\Gamma$-condition is known to hold for the setting of exponential utility optimization~\cite{Morlais}, it is 
rather restrictive, and in fact, stronger than the local Lipschitz continuity. Furthermore, in the anticipated settings,
the comparison principle does not hold generally even when the $A_\Gamma$-condition is satisfied.
Although the fixed point approach~\cite{Cohen-Elliott, Kazi} does not rely on the comparison principle 
at least for small terminal values, it requires the second-order differentiability of the driver
which is difficult to establish in the presence of the general path dependence.

In this paper, we firstly extend the quadratic-exponential structure condition of \cite{Barrieu-ElKaroui, ElKaroui-Matoussi}
to allow the dependence on $Y$'s future paths,  and then derive the universal bounds on $(Y,Z,\psi)$
under a general bounded terminal condition.
This bounds are then used to prove a stability result under a general non-Markovian setting.
Under the Markovian setting, this stability result leads to the compactness result
for the deterministic map defined by $u(t,x)=Y^{t,x}_t$, which then allows us to prove the existence of the solution
in the absence of the $A_\Gamma$-condition. It also provides some regularities on $(Z,\psi)$ with respect to the 
forward process.  As a by product, it makes the $A_\Gamma$-condition unnecessary for the existence, uniqueness
and Malliavin's differentiability of quadratic-exponential growth (non-anticipated) BSDEs 
under the Markovian setting studied in Section 6 of \cite{FT-Qexp}.
For a non-Markovian setting, we reintroduce the $A_\Gamma$-condition and make use of 
our previous result in \cite{FT-Qexp} to prove the existence of the unique solution.
We also give a sufficient condition for the comparison principle to hold.

%%%%%%%%%%%%%%%%%%%%%%%%%%%%%%%%%%%
\section{Preliminaries}
\label{sec-Preliminaries}
%%%%%%%%%%%%%%%%%%%%%%%%%%%%%%%%%%%
\subsection{General Setting}
%%%%%%%%%%%%%%%%%%%%%%%%%%%%%%%%%%
Let us first state the general setting to be used  throughout the paper.
$T>0$ is some bounded time horizon.
The space $(\Omega_W,\calf_W,\mbb{P}_W)$
is the usual canonical space for a $d$-dimensional Brownian motion equipped with the Wiener measure $\mbb{P}_W$.
We also denote  $(\Omega_\mu,\calf_\mu,\mbb{P}_\mu)$ as a product of 
canonical spaces $\Omega_\mu:=\Omega_\mu^1\times \cdots\times \Omega_\mu^k$,
$\calf_\mu:=\calf_\mu^1\times \cdots \times \calf_\mu^k$ and $\mbb{P}_\mu^1\times \cdots \times \mbb{P}_\mu^k$ with some 
constant $k\in\mbb{N}$,
on which each $\mu^i$ is a Poisson measure with a compensator $\nu^i(de)dt$. Here, $\nu^i(de)$ is a $\sigma$-finite 
measure on $\mbb{R}_0=\mbb{R}\backslash\{0\}$ satisfying $\int_{\mbb{R}_0} |e|^2 \nu^i(de)<\infty$.
For notational simplicity, we write $(E,\cale):=(\mbb{R}^k_0,\calb(\mbb{R}_0)^k)$.
Throughout the paper, we work on the filtered probability space 
$(\Omega, \calf, \mbb{F}=(\calf_t)_{t\in[0,T]}, \mbb{P})$, where the  space $(\Omega,\calf,\mbb{P})$
is the product of the canonical spaces $(\Omega_W\times  \Omega_\mu, \calf_W\times\calf_\mu,\mbb{P}_W\times \mbb{P}_\mu)$,
and that the filtration $\mbb{F}=(\calf_t)_{t\in[0,T]}$ is the 
canonical filtration completed for $\mbb{P}$ and satisfying the usual conditions.
In this construction, $(W,\mu^1,\cdots,\mu^k)$ are independent.
We use a vector notation $\mu(\omega, dt,de):=(\mu^1(\omega, dt,de^1), \cdots, \mu^k(\omega, dt,de^k))$
and denote the compensated Poisson measure as $\wt{\mu}:=\mu-\nu$.
$\mbb{F}$-predictable $\sigma$-field on $\Omega\times [0,T]$ is denoted by $\calp$.
It is well-known that the weak property of predictable representation holds in this setup~(see for example \cite{He-Wang-Yang} chapter XIII).

%\begin{remark}
%Under construction....
%\end{remark}
%%%%%%%%%%%%%%%%%%%%%%%%%%%%%%%%%%%%
\subsection{Notation}
%%%%%%%%%%%%%%%%%%%%%%%%%%%%%%%%%%%%
We denote a generic constant by $C$ which may change line by line. 
We write $C=C(a,b,c,\cdots)$ when the constant depends only on the parameters $(a,b,c,\cdots)$.
$\calt^t_s$ denotes
the set of $\mbb{F}$-stopping times $\tau:\Omega \rightarrow [s,t]$. 
We denote the conditional expectation with respect to $\calf_t$ by $\mbb{E}_{\calf_t}[\cdot]$ or $\mbb{E}[\cdot|\calf_t]$.
Under a probability measure $\mbb{Q}$ different from $\mbb{P}$, we explicitly denote it, for example, by $\mbb{E}^{\mbb{Q}}_{\calf_t}[\cdot]$.
Sometimes we use the abbreviations 
$||x||_{[s,t]}:=\sup_{v\in[s,t]}|x_v|$ and $\Theta_v:=(Y_v,Z_v,\psi_v)$.

We introduce the following spaces.  $p\in \mbb{N}$ is assumed to be $p\geq 2$.\\
\bull $\mbb{D}[s,t]$ is the set of real valued \cadlag functions $(q_v)_{v\in[s,t]}$.\\
\bull $\mbb{S}^p[s,t]$ is the set of real (or vector) valued \cadlag $\mbb{F}$-adapted processes $(X_v)_{v\in[s,t]}$ such that
\be
||X||_{\mbb{S}^p[s,t]}:=\mbb{E}\bigl[ \sup_{v\in[s,t]}|X_v|^p\bigr]^\frac{1}{p}<\infty.\nn
\ee
\bull $\mbb{S}^\infty[s,t]$ is the set of real (or vector) valued \cadlag $\mbb{F}$-adapted processes $(X_v)_{v\in[s,t]}$ 
which are essentially bounded, i.e.
\be
||X||_{\mbb{S}^\infty[s,t]}:=\bigl|\bigl| \sup_{v\in[s,t]} |X_v|\bigr|\bigr|_{\infty}<\infty.\nn
\ee
Here, $||x||_{\infty}:=\inf\bigl\{c\in \mbb{R}~;~\mbb{P}(\{|x|\leq c\})=1 \bigr\}$.\\
\bull $\mbb{H}^p[s,t]$ is the set of progressively measurable real (or vector) valued processes $(Z_v)_{v\in[s,t]}$
such that
\be
||Z||_{\mbb{H}^p[s,t]}:=\mbb{E}\Bigl[\Bigl(\int_s^t |Z_v|^2 dv\Bigr)^\frac{p}{2}\Bigr]^\frac{1}{p}<\infty.\nn
\ee
\bull $\mbb{L}^2(E,\nu)$ (or simply $\mbb{L}^2(\nu)$) is the set of $k$-dimensional vector-valued functions $\psi=(\psi^i)_{1\leq i\leq k}$
for which the each component $\psi^i:\mbb{R}_0\rightarrow \mbb{R}$ is $\calb(\mbb{R}^0)$-measurable and
\be
||\psi||_{\mbb{L}^2(E,\nu)}:=\Bigl(\sum_{i=1}^k\int_{\mbb{R}_0}|\psi^i(e)|^2\nu^i(de)\Bigr)^\frac{1}{2}<\infty.\nn
\ee
\bull $\mbb{L}^\infty(E,\nu)$ (or simply $\mbb{L}^\infty(\nu)$) is the set of functions $\psi=(\psi^i)_{1\leq i\leq k}$ 
for which the each component $\psi^i:\mbb{R}_0\rightarrow \mbb{R}$ is $\calb(\mbb{R}^0)$-measurable and bounded 
$\nu^i(de)$-a.e. with the standard essential supremum norm.
\\
\bull $\mbb{J}^p[s,t]$ is the set of functions $\psi=(\psi^i)_{1\leq i\leq k}$ with 
$\psi^i:\Omega\times [s,t]\times \mbb{R}_0\rightarrow \mbb{R}$ being $\calp\otimes \calb(\mbb{R}_0)$-measurable
(or we simply say $\psi$ is $\calp\otimes \cale$-measurable) and satisfy
\be
||\psi||_{\mbb{J}^p[s,t]}:=\mbb{E}\Bigl[\Bigl(\sum_{i=1}^k\int_s^t\int_{\mbb{R}_0}|\psi^i_v(e)|^2\nu^i(de)dv\Bigr)^\frac{p}{2}
\Bigr]^\frac{1}{p}<\infty.\nn
\ee
\bull We denote $\calk^p[s,t]=\mbb{S}^p[s,t]\times \mbb{H}^p[s,t]\times \mbb{J}^p[s,t]$ with the norm
\be
||(Y,Z,\psi)||_{\calk^p[s,t]}:= ||Y||_{\mbb{S}^p[s,t]}+||Z||_{\mbb{H}^p[s,t]}+||\psi||_{\mbb{J}^p[s,t]}.\nn
\ee
For notational simplicity, hereafter we write
\bea
\int_s^t\int_E \psi_r(e)\wt{\mu}(dr,de):=\sum_{i=1}^k\int_s^t\int_{\mbb{R}_0}\psi^i_r(e)\wt{\mu}^i(dr,de)\nn
\eea
and use similar abbreviations for the  integrations with respect to $(\mu,\nu)=(\mu^i,\nu^i)_{1\leq i\leq k}$.
\\
\\
%%%%%%%%%%%%%%%%%%%%%%%%%%%%%%%%%%%%%%%
\bull $\mbb{J}^\infty[s,t]$ is the set of $\calp\otimes \cale$-measurable functions $\psi=(\psi^i)_{1\leq i\leq k}$
essentially bounded with respect to the measure $d\mbb{P}\otimes\nu(de)\otimes dt$ i.e. 
\be
||\psi||_{\mbb{J}^\infty[s,t]}:=\bigl|\bigl|\underset{v\in[s,t]}{{\rm ess~sup}}||\psi_v||_{\mbb{L}^\infty(E,\nu)}\bigr|\bigr|_{\infty}<\infty.\nn
\ee
%%%%%%%%%%%%%%%%%%%%%
\bull $\mbb{H}^2_{BMO}[s,t]$ is the set of real (or vector) valued progressively measurable processes $(Z_v)_{v\in[s,t]}$ such that
\be
||Z||^2_{\mbb{H}^2_{BMO}[s,t]}:=\sup_{\tau\in\calt^t_s}\Bigr|\Bigr|\mbb{E}_{\calf_\tau}\Bigl[\int_\tau^t|Z_r|^2 dr\Bigr]
\Bigr|\Bigr|_{\infty}<\infty. \nn
\ee
\bull $\mbb{J}^2_{B}[s,t]$ is the set of $\calp\otimes \cale$-measurable functions such that
\be
||\psi||_{\mbb{J}^2_{B}[s,t]}^2:=\sup_{\tau\in\calt^t_s}\Bigr|\Bigr|\mbb{E}_{\calf_\tau}\Bigl[
\int_\tau^t\int_E |\psi_r(e)|^2 \nu(de)dr\Bigr]\Bigr|\Bigr|_{\infty}<\infty.\nn
\ee
\bull $\mbb{J}^2_{BMO}[s,t]$ is the set of  $\calp\otimes \cale$-measurable functions such that
\be
||\psi||_{\mbb{J}^2_{BMO}[s,t]}^2:=\sup_{\tau\in\calt^t_s}\Bigr|\Bigr|\mbb{E}_{\calf_\tau}\Bigl[
\int_\tau^t\int_E |\psi_r(e)|^2 \nu(de)dr\Bigr]+(\Delta M_\tau)^2\Bigr|\Bigr|_{\infty}<\infty,\nn
\ee
where $\Del M_\tau:=\int_E\psi_\tau(e)\mu(\{\tau\},de)$. 
See Section 2.3 of \cite{FT-Qexp} and references therein for the details of BMO-martingales with jumps.
We frequently omit $[s,t]$ if it is obvious from the context.

%%%%%%%%%%%%%%%%%%%%%%%%%%%%%%%%%%%%%%%%%%%%%%%%%%%%%%%%%%%%%%%%%%%%%%%%%%%%%%%%%%%%%%%%%%%%%%%%%%
\subsection{Some relations among the jump norms $\mbb{J}^\infty, \mbb{J}^2_B, \mbb{J}^2_{BMO}$ }
%%%%%%%%%%%%%%%%%%%%%%%%%%%%%%%%%%%%%%%%%%%%%%%%%%%%%%%%%%%%%%%%%%%%%%%%%%%%%%%%%%%%%%%%%%%%%%%%%%
By a simple adaptation of Corollary 1 in \cite{Morlais-old}, we get the next Lemma.

%%%%%%%%%%%%%%%%%%%%%%%%%%%%%%%%%
\begin{lemma}
\label{lemma-J-bound}
Let $\psi$ be in $\mbb{J}^2[0,T]$, and define a square-integrable pure jump martingale $(M_t)_{t\in[0,T]}$
by $M_t:=\int_0^t \int_E \psi_s(e)\wt{\mu}(ds,de)$. The jump  $\Del M$ at time $t$ is given by
\bea
\Del M_t:=M_t-M_{t-}=\int_E \psi_t(e)\mu(\{t\},de)~.
\label{ap-eq-delM}
\eea
Then the following two conditions are equivalent:\\
(1) $||\psi||_{\mbb{J}^\infty[0,T]}$ is finite. \\
(2) $\sup_{\tau \in \calt_0^T}||\Del M_\tau||_{\infty}$ is finite.\\
Moreover, the above two quantities coincide when they exist i.e.
\be
||\psi||_{\mbb{J}^\infty}=\sup_{\tau\in\calt_0^T}||\Del M_\tau||_{\infty}~.\nn
\ee
\begin{proof}
$(1)\Rightarrow (2)$. Let assume $\psi\in \mbb{J}^2[0,T]\cap\mbb{J}^\infty[0,T]$.
By construction, only the jump times of the Poisson measure $\mu(dt,de)$ contributes $|\Del M|$.
Since the density of $\mu(dt,de)$ is given by $\nu(de)\otimes dt$, it is obvious to see
$|\Del M_{\tau}|\leq ||\psi||_{\mbb{J}^\infty[0,T]}$ a.s. for every stopping time $\tau\in\calt^T_0$
and hence $\sup_{\tau\in\calt_0^T}||\Del M_\tau||_{\infty}\leq ||\psi||_{\mbb{J}^\infty[0,T]}$.

$(2)\Rightarrow(1)$. Assume conversely, $C:=\sup_{\tau \in \calt_0^T}||\Del M_\tau||_{\infty}<\infty$.
By (\ref{ap-eq-delM}), one sees
\be
|\psi_\tau(e)|\leq C\qquad {\rm a.s.} 
\label{psi-constraint}
\ee
for every pair $(\tau,e)$ of the jump time $(\tau)$ and its
associated mark $(e)$ of the random measure $\mu(dt,de)$.
Let us define a new process $\overline{\psi}\in \mbb{J}^2[0,T]\cap\mbb{J}^\infty[0,T]$ by the next truncation:
\bea
\bar{\psi}_t(e):=\psi_t(e)\bold{1}_{\{|\psi_t(e)|\leq C\}}~\quad \forall (\omega, t,e)\in \Omega \times [0,T]\times E.\nn
\eea
Notice that $\psi$ and $\overline{\psi}$ are equal a.s. on every jump time and its associated mark of $\mu$.
As a consequence, one has
\bea
0&=&\mbb{E}\Bigl[\int_0^T \int_E |\psi_t(e)-\overline{\psi}_t(e)|^2\mu(dt,de)\Bigr]\nn \\
&=&\mbb{E}\Bigl[\int_0^T \int_E |\psi_t(e)-\overline{\psi}_t(e)|^2\nu(de)dt\Bigr]\nn.
\eea
This means $\psi=\overline{\psi}$ in $\mbb{J}^2[0,T]$ and hence, in particular, 
$d\mbb{P}\otimes \nu(de)\otimes dt$-a.e.
Therefore, $||\psi||_{\mbb{J}^\infty[0,T]}=||\overline{\psi}||_{\mbb{J}^\infty[0,T]}\leq C$
and (1) holds. This establishes the equivalent of (1) and (2).
Combining the two results, one can conclude 
$||\psi||_{\mbb{J}^\infty[0,T]}=\sup_{\tau\in\calt_0^T}||\Del M_\tau||_{\infty}$.
\end{proof}
\end{lemma}
%%%%%%%%%%%%%%%%%%%%%%%
\begin{remark}
Note that $\psi$ must be a predictable process. This fact makes the constraint only at the jump points in (\ref{psi-constraint}) 
be translated into the whole domain of $\psi$. \end{remark}
%%%%%%%%%%%%%%%%%%%%%%%%%%%%%%%%%%%%%%%%%
Using the above result, one obtains the following relation among the different jump norms.
\begin{lemma}
The following two conditions are equivalent: \\
(1) $\psi\in \mbb{J}^2_{B}[0,T]\cap \mbb{J}^\infty[0,T]$~.\\
(2) $\psi\in \mbb{J}^2_{BMO}[0,T]$~.\\
Moreover, the following inequality holds:
\bea
(||\psi||^2_{\mbb{J}^2_B[0,T]}\vee ||\psi||^2_{\mbb{J}^\infty[0,T]})\leq ||\psi||^2_{\mbb{J}^2_{BMO}[0,T]}
\leq ||\psi||^2_{\mbb{J}^2_B[0,T]}+||\psi||^2_{\mbb{J}^\infty[0,T]}.\nn
\eea
\begin{proof}
$(1)\Rightarrow (2)$. Let $\psi$ be in $\mbb{J}^2_B[0,T]\cap \mbb{J}^\infty[0,T]$. Since $\mbb{J}^2_B \subset\mbb{J}^2$,
Lemma~\ref{lemma-J-bound} implies that
$||\psi||_{\mbb{J}^\infty[0,T]}=\sup_{\tau\in\calt_0^T}||\Del M_\tau||_{\infty}$
where $\Del M_{\tau}:=\int_E \psi_{\tau}(e)\mu(\{\tau\},de)~$ as defined before.
One then obtains
\bea
||\psi||^2_{\mbb{J}^2_{BMO}[0,T]}&=&\sup_{\tau\in\calt_0^T}\Bigl|\Bigl| \mbb{E}_{\calf_\tau}\Bigl[
\int_\tau^T\int_E |\psi_r(e)|^2 \nu(de)dr\Bigr]+(\Delta M_\tau)^2\Bigr|\Bigr|_{\infty}\nn \\
&\leq &||\psi||_{\mbb{J}^2_B[0,T]}^2+\sup_{\tau\in\calt_0^T}||\Del M_\tau||_{\infty}^2
=||\psi||^2_{\mbb{J}^2_B[0,T]}+||\psi||_{\mbb{J}^\infty[0,T]}^2~. \nn
\eea
Thus (2) holds.

$(2)\Rightarrow (1)$. On the other hand let $\psi\in\mbb{J}^2_{BMO}[0,T]$.  By definition of $\mbb{J}^2_{BMO}$-norm, 
one has
\bea
\sup_{\tau\in\calt_0^T}\Bigl|\Bigl|\mbb{E}_{\calf_\tau}\Bigl[\int_\tau^T |\psi_r(e)|^2 \nu(de)dr\Bigr]\Bigr|\Bigr|_{\infty}\vee
\sup_{\tau\in\calt_0^T}||\Del M_\tau||^2_{\infty}\leq ||\psi||_{\mbb{J}^2_{BMO}}^2~.\nn
\eea
Since $\mbb{J}^2_{BMO}\subset \mbb{J}^2_B$, Lemma~\ref{lemma-J-bound} once again implies
\bea
||\psi||_{\mbb{J}^2_{B[0,T]}}^2 \vee ||\psi||_{\mbb{J}^\infty[0,T]}^2\leq ||\psi||^2_{\mbb{J}^2_{BMO}[0,T]}~~. \nn
\eea
The last claim direct follows from the above two inequalities.
\end{proof}
\end{lemma}

\begin{remark}
\label{remark-essential-bound}
When $\psi$ is given as a part of BSDE solution $(Y,Z,\psi)$ as in (\ref{eq-qg-bsde-nonM}),
$\psi$ can be defined only up $d\mbb{P}\otimes \nu(de)\otimes dt$-a.e.
Thus, if one has a $\psi\in \mbb{J}^\infty$, one can freely work on its version 
$(\widetilde{\psi}_t(\omega, e))_{(\omega,t,e)\in\Omega\times [0,T]\times E}$  which is everywhere bounded
(as in $\bar{\psi}$ used in the proof of Lemma~\ref{lemma-J-bound}).
This fact is being used in some existing literature.
\end{remark}

%%%%%%%%%%%%%%%%%%%%%%%%%%%%%%%%%%%%%%%%%%%%%%%%%%%%%%%%%%%%%%%%%%%%%%%
\section{A priori estimates}
%%%%%%%%%%%%%%%%%%%%%%%%%%%%%%%%%%%%%%%%%%%%%%%%%%%%%%%%%%%%%%%%%%%%%%%
\subsection{Universal bounds}
In this section, we consider various a priori estimates regarding anticipated quadratic-exponential growth 
BSDEs with jumps in a general non-Markovian setup.
We are interested in the following ABSDE for $t\in[0,T]$:
\bea
Y_t=\xi+\int_t^T \mbb{E}_{\calf_r}f\Bigl(r,(Y_v)_{v\in[r,T]},Y_r,Z_r,\psi_r\Bigr)dr
-\int_t^T Z_r dW_r-\int_t^T\int_E \psi_r(e)\wt{\mu}(dr,de), 
\label{eq-qg-bsde-nonM}
\eea
where 
$f:\Omega\times [0,T]\times \mbb{D}[0,T]\times \mbb{R}\times \mbb{R}^{1\times d}\times \mbb{L}^2(E,\nu)\rightarrow \mbb{R}$,
and $\xi$ is an $\calf_T$-measurable random variable.

\begin{assumption}
\label{assumption-structure-nonM}
(i) The driver $f$ is a map such that for every $(y,z,\psi)\in\mbb{R}\times \mbb{R}^{1\times d}\times \mbb{L}^2(E,\nu)$
and any \cadlag $\mbb{F}$-adapted process $(Y_v)_{v\in[0,T]}$, the process $\bigl(\mbb{E}_{\calf_t}f(t,(Y_v)_{v\in[t,T]},y,z,\psi), t\in[0,T]\bigr)$ is $\mbb{F}$-progressively measurable, and the map $(y,z,\psi)\rightarrow f(\cdot,y,z,\psi)$
is continuous. \\
(ii) For every $(q,y,z,\psi)\in\mbb{D}[0,T]\times \mbb{R}\times \mbb{R}^{1\times d}\times \mbb{L}^2(E,\nu)$,
there exist constants $\beta,\del\geq 0$, $\gamma>0$ and a positive progressively measurable process $(l_v,v\in[0,T])$ such that
\bea
&&-l_t-\del\bigl(\sup_{v\in[t,T]}|q_v|\bigr)-\beta|y|-\frac{\gamma}{2}|z|^2-\int_E j_\gamma (-\psi(e))\nu(de) ~\leq~ f\bigl(t,(q_v)_{v\in[t,T]},y,z,\psi\bigr)\nn \\
&&\qquad \leq l_t+\del\bigl(\sup_{v\in[t,T]}|q_v|\bigr)+\beta|y|+\frac{\gamma}{2}|z|^2
+\int_E j_\gamma (\psi(e))\nu(de) \nn
\eea
$d\mbb{P}\otimes dt$-a.e. $(\omega,t)\in\Omega\times [0,T]$, where
$j_\gamma(u):=\frac{1}{\gamma}(e^{\gamma u}-1-\gamma u)$.
\\
(iii) $||\xi||_{\infty}, ||l||_{\mbb{S}^\infty}<\infty$.
\end{assumption}

\begin{lemma}
\label{lemma-z-psi-bound}
Under Assumption~\ref{assumption-structure-nonM}, if there exists a bounded solution $(Y,Z,\psi)\in \mbb{S}^\infty\times \mbb{H}^2\times 
\mbb{J}^2$ to the ABSDE (\ref{eq-qg-bsde-nonM}), then $Z\in \mbb{H}^2_{BMO}$ and $\psi\in \mbb{J}^2_{BMO}$ (hence $\psi\in \mbb{J}^\infty$)
and they satisfy
\bea
&&||Z||^2_{\mbb{H}^2_{BMO}}\leq \frac{e^{4\gamma ||Y||_{\mbb{S}^\infty}}}{\gamma^2}\Bigl(1+2\gamma T\bigl[||l||_{\mbb{S}^\infty}
+(\beta+\del)||Y||_{\mbb{S}^\infty}\bigr]\Bigr), \nn \\
&&||\psi||^2_{\mbb{J}^2_{BMO}}\leq \frac{e^{4\gamma ||Y||_{\mbb{S}^\infty}}}{\gamma^2}\Bigl(2+4\gamma T\bigl[||l||_{\mbb{S}^\infty}
+(\beta+\del)||Y||_{\mbb{S}^\infty}\bigr]\Bigr)+4||Y||^2_{\mbb{S}^\infty}.\nn
\eea
\begin{proof}
It follows from Lemma~3.1 \cite{FT-Qexp} by a simple replacement of $||l||_{\mbb{S}^\infty}$ with $||l||_{\mbb{S}^\infty}+
\del ||Y||_{\mbb{S}^\infty}$. One also needs the fact that 
$||\psi||_{\mbb{J}^2_{BMO}}^2\leq ||\psi||_{\mbb{J}^2_B}^2+||\psi||_{\mbb{J}^\infty}^2$ and 
$||\psi||_{\mbb{J}^\infty}\leq 2||Y||_{\mbb{S}^\infty}$ from Lemma~\ref{lemma-J-bound}.
We give details in Appendix~\ref{ap-L3-1}.
\end{proof}
\end{lemma}

\begin{lemma}
\label{lemma-y-bound}
Under Assumption~\ref{assumption-structure-nonM}, if there exists a bounded solution $(Y,Z,\psi)\in \mbb{S}^\infty\times \mbb{H}^2\times 
\mbb{J}^2$ to the ABSDE (\ref{eq-qg-bsde-nonM}), then $Y$ has the following estimate
\be
||Y||_{\mbb{S}^\infty}\leq \exp\Bigl(T\bigl(\beta+\del e^{\beta T}\bigr)\Bigr)\bigl(||\xi||_{\infty}+T||l||_{\mbb{S}^\infty}\bigr)~.\nn
\ee
\begin{proof}
Applying Mayer-Ito formula, one obtains
\bea
d(e^{\beta s}|Y_s|)&=&e^{\beta s}\Bigl(\beta |Y_s|ds+\sign(Y_{s-})dY_s+dL_s\Bigr). \nn 
\eea 
Here, $(L_s)_{s\in[0,T]}$ is a non-decreasing process including a local time $L^c$ as
\be
dL_s=dL_s^c+\int_E\Bigl(|Y_{s-}+\psi_s(e)|-|Y_{s-}|-\sign(Y_{s-})\psi_s(e)\Bigr)\mu(ds,de)~.\nn
\ee
Note that 
\be
|y+\psi|-|y|-\sign(y)\psi=|y+\psi|-\sign(y)(y+\psi)\geq 0.
\label{eq-ineq}
\ee
Let us introduce the following processes $(B_s)_{s\in[0,T]}$ and $(C_s)_{s\in[0,T]}$ by
\bea
dB_s&=&-\sign(Y_s)\mbb{E}_{\calf_s}f\bigl(s,(Y_v)_{v\in[s,T]},\Theta_s)ds\nn \\
&&+\Bigl(l_s+\del \mbb{E}_{\calf_s}\bigl(\sup_{v\in[s,T]}|Y_v|\bigr)+\beta|Y_s|+\frac{\gamma}{2}|Z_s|^2
+\int_E j_\gamma(\sign(Y_s)\psi_s(e))\nu(de)\Bigr)ds~,\nn \\
dC_s&=&e^{\beta s}(dB_s+dL_s)+\frac{\gamma}{2}(e^{2\beta s}-e^{\beta s})|Z_s|^2 ds\nn \\
&&+\int_E \Bigl(j_\gamma(e^{\beta s}\sign(Y_s)\psi_s(e))-e^{\beta s}j_\gamma(\sign(Y_s)\psi_s(e))\Bigr)\nu(de)ds~.\nn
\eea
Note that both of $B$ and $C$ are non-decreasing processes. As for the process $B$, this follows from Assumption~\ref{assumption-structure-nonM}.
As for the process $C$,  it follows from the fact that $k\geq 1$, $j_\gamma (ku)-kj_\gamma(u)=\frac{1}{\gamma}(e^{k\gamma u}-ke^{\gamma u}-1+k)\geq 0$,
which makes the last line positive. One then sees
\bea
&&d\Bigl(e^{\beta s}|Y_s|+\int_0^s e^{\beta r}\bigl(l_r+\del \mbb{E}_{\calf_r}(\sup_{v\in[r,T]}|Y_v|)\bigr)dr\Bigr)=e^{\beta s}\sign(Y_{s-})\Bigl(Z_s dW_s+\int_E \psi_s(e)\wt{\mu}(ds,de)\Bigr)\nn\\
&&\hspace{25mm}-\int_E j_\gamma\bigl(e^{\beta s}\sign(Y_s)\psi_s(e)\bigr)\nu(de)ds-\frac{\gamma}{2}|e^{\beta s}\sign(Y_s)Z_s|^2 ds+dC_s.
\label{eq-d-group}
\eea

We now investigate the process $P_t,t\in[0,T]$ defined by
\bea
P_t:=\exp\Bigl(\gamma e^{\beta t}|Y_t|+\gamma \int_0^t e^{\beta r}\bigl(l_r+\del \mbb{E}_{\calf_r}(\sup_{v\in[r,T]}|Y_v|)
\bigr)dr\Bigr),\quad t\in[0,T],\nn
\eea
where $P\in \mbb{S}^\infty$ is clearly seen. Applying Ito formula, one obtains that 
\bea
&&dP_t=P_{t-}\gamma d\Bigl(e^{\beta t}|Y_t|+\int_0^t e^{\beta r}\bigl(l_r+\del \mbb{E}_{\calf_r}\sup_{v\in[r,T]}|Y_v|\bigr)dr\Bigr)
+P_t\frac{\gamma^2}{2}|e^{\beta t}\sign(Y_t)Z_t|^2 dt\nn \\
&&\qquad +P_{t-}\int_E \Bigl(e^{\gamma e^{\beta t}(|Y_{t-}+\psi_t(e)|-|Y_{t-}|)}-1-\gamma e^{\beta t}\sign(Y_{t-})\psi_t(e)\Bigr)\mu(dt,de)\nn \\
&&=P_{t-}\Bigl(\gamma e^{\beta t}\sign(Y_t)Z_t dW_t+\int_E \Bigl(\exp\bigl(\gamma e^{\beta t}
\sign(Y_{t-})\psi_t(e)\bigr)-1\Bigr)\wt{\mu}(dt,de)+dC_t^\prime\Bigr)
\label{eq-dP}
\eea
where $(C^\prime_s)_{s\in[0,T]}$ is another non-decreasing (see (\ref{eq-ineq})) process defined by
\be
dC^\prime_t=\gamma dC_t+\int_E\Bigl(e^{\gamma e^{\beta t}(|Y_{t-}+\psi_t(e)|-|Y_{t-}|)}-e^{\gamma e^{\beta t}\sign(Y_{t-})\psi_t(e)}
\Bigr)\mu(dt,de).
\label{eq-dCprime}
\ee
The details of the derivation of (\ref{eq-dP}) are given in Appendix~\ref{ap-L3-2}.

Since $(P,Y,Z,\psi)\in \mbb{S}^\infty\times \mbb{S}^\infty\times \mbb{H}^2_{BMO}\times \mbb{J}^2_{BMO}$, one sees the process $P$
is a true submartingale. Therefore, it follows that, for any $t\in[0,T]$, 
\bea
\exp\Bigl(\gamma e^{\beta t}|Y_t|\Bigr)&\leq& \mbb{E}_{\calf_t}\left[
\exp\Bigl(\gamma e^{\beta T}|\xi|+\gamma \int_t^T e^{\beta r}\bigl(l_r+\del \mbb{E}_{\calf_r}(\sup_{v\in[r,T]}|Y_v|)\bigr)dr\Bigr)
\right]\nn\\
&\leq &\exp\Bigl(\gamma e^{\beta T}(||\xi||_{\infty}+T||l||_{\mbb{S}^\infty})+\gamma \del e^{\beta T}
\int_t^T ||Y||_{\mbb{S}^\infty[r,T]}~dr\Bigr)~{\rm a.s.} \nn 
\eea
Thus, 
$\displaystyle |Y_t|\leq e^{\beta T}(||\xi||_{\infty}+T||l||_{\mbb{S}^\infty})+\del e^{\beta T}\int_t^T ||Y||_{\mbb{S}^\infty[r,T]}~dr~{\rm a.s.}$ 
Since the right-hand side is non-increasing in $t$, 
the same inequality holds with the left-hand side replaced by $\sup_{s\in[t,T]}|Y_s|$. Hence equivalently,
\bea
||Y||_{\mbb{S}^\infty[t,T]}\leq e^{\beta T}(||\xi||_{\infty}+T||l||_{\mbb{S}^\infty})+\del e^{\beta T}\int_t^T 
||Y||_{\mbb{S}^\infty[r,T]}~dr~.\nn
\eea
Now using the backward Gronwall inequality~\footnote{See, for example, Corollary 6.61~\cite{Pardoux-Rascanu}}, one
obtains the desired result. 
\end{proof}
\end{lemma}

\begin{definition}
We define the set of parameters $A:=(||\xi||_{\infty},||l||_{\mbb{S}^\infty}, \del, \beta, \gamma, T)$
which control the universal bounds on $(||Y||_{\mbb{S}^\infty}, ||Z||_{\mbb{H}^2_{BMO}}, ||\psi||_{\mbb{J}^2_{BMO}})$.
\end{definition}

As a result of Lemmas~\ref{lemma-z-psi-bound} and \ref{lemma-y-bound}, one sees
the norms of $||Y||_{\mbb{S}^\infty}, ||Z||_{\mbb{H}^2_{BMO}}, ||\psi||_{\mbb{J}^2_{BMO}}$
are solely controlled by the set of parameters in $A$.
In the next subsection, we introduce the local Lipschitz continuity.

%%%%%%%%%%%%%%%%%%%%%%%%%%%%%%%%%%%%%%%%%%%%%%%%%%
\subsection{Stability and Uniqueness}
%%%%%%%%%%%%%%%%%%%%%%%%%%%%%%%%%%%%%%%%%%%%%%%%%%
\begin{assumption}
\label{assumption-LLC-nonM}
For each $M>0$, and for every $(q,y,z,\psi),~(q^\prime,y^\p,z^\p,\psi^\p)\in \mbb{D}[0,T]\times \mbb{R}\times \mbb{R}^{1\times d}
\times \mbb{L}^2(E,\nu)$ satisfying 
$\sup_{v\in[0,T]}|q_v|, \sup_{v\in[0,T]}|q^{\p}_v|, ~|y|, ~|y^\p|, ~||\psi||_{\mbb{L}^\infty(\nu)}, ~||\psi^\p||_{\mbb{L}^\infty(\nu)}
\leq M$,  there exists some positive constant $K_M$ (depending on $M$) such that
\bea
&&|f(t,(q_v)_{v\in[t,T]},y,z,\psi)-f(t,(q^\p_v)_{v\in[t,T]},y^\p,z^\p,\psi^\p)|\nn \\
&&\qquad \leq K_M\Bigl(\sup_{v\in[t,T]}|q_v-q^\p_v|+|y-y^\p|+||\psi-\psi^\p||_{\mbb{L}^2(\nu)}\Bigr)\nn \\
&&\qquad +K_M\bigl(1+|z|+|z^\p|+||\psi||_{\mbb{L}^2(\nu)}+||\psi^\p||_{\mbb{L}^2(\nu)}\bigr)|z-z^\p| 
\label{eq-LLC-nonM}
\eea
$d\mbb{P}\otimes dt$-a.e. $(\omega,t)\in\Omega\times [0,T]$.
\end{assumption}
\begin{remark}
Instead of directly making the driver $f$ path-dependent, one can include the conditional
expectations such as $\mbb{E}_{\calf_t}(Y_{t+\del}), \mbb{E}_{\calf_t} (\int_t^T Y_s ds)$ 
as done in \cite{Oksendal}.  In this work, we adopt the former approach since it allows 
the general dependence without specifying a concrete form.
\end{remark}

Let us introduce the two ABSDEs for $t\in[0,T]$,  with $i=\{1,2\}$, 
\bea
Y^i_t=\xi^i+\int_t^T \mbb{E}_{\calf_r}f^i\Bigl(r,(Y^i_v)_{v\in[r,T]},Y^i_r,Z^i_r,\psi^i_r\Bigr)dr-
\int_t^T Z_r^i dW_r-\int_t^T\int_E \psi_r^i(e)\wt{\mu}(dr,de).
\label{eq-yi-stability}
\eea
Let us put
$\del Y:=Y^1-Y^2, \del Z:=Z^1-Z^2, \del \psi:=\psi^1-\psi^2$, and 
\be
\del f(r):=(f^1-f^2)\bigl(r,(Y^1_v)_{v\in[r,T]}, Y^1_r,Z^1_r,\psi^1_r\bigr)~.\nn
\ee
Then, we have the following stability result.

\begin{proposition}
\label{prop-stability}
Suppose that the data $(\xi^i,f^i)_{1\leq i\leq 2}$ satisfy Assumptions~\ref{assumption-structure-nonM}
and \ref{assumption-LLC-nonM}. If the two ABSDEs (\ref{eq-yi-stability}) have bounded solutions $(Y^i,Z^i,\psi^i)_{1\leq i\leq 2}\in 
\mbb{S}^\infty\times \mbb{H}^2\times \mbb{J}^2$,  then for any $p> 2q_{*}$ 
\be
||\del Y||_{\mbb{S}^p[0,T]}\leq C_1\mbb{E}\Bigl[|\del\xi|^p+\Bigl(\int_0^T\mbb{E}_{\calf_r}|\del f(r)|dr\Bigr)^p\Bigr]^\frac{1}{p}
\label{eq-y-stability}
\ee
and for any $p\geq 2, \bq\geq q_*$ 
\bea
&&\bigl|\bigl|(\del Y,\del Z,\del\psi)\bigr|\bigr|_{\calk^p[0,T]}\leq C_2\mbb{E}\Bigl[|\del \xi|^{p\bq^2}+\Bigl(\int_0^T \mbb{E}_{\calf_r}|\del f(r)|dr\Bigr)^{p\bq^2}\Bigr]^\frac{1}{p\bq^2}
\label{eq-z-psi-stability}
\eea
where $q_*\in(1,\infty)$ is a constant depending only on $(K_\cdot, A)$, $C_1=C(p,K_\cdot,A)$ and $C_2=(p,\bq,K_\cdot,A)$ are two positive constants.
\begin{proof}
Note that one can apply (\ref{eq-LLC-nonM}) globally with fixed $K_M$ by choosing 
$M$ larger than the bounds implied from Lemmas~\ref{lemma-z-psi-bound} and \ref{lemma-y-bound}.
Let fix such an $M$ in the remainder.
Define the $\mbb{R}^d$-valued progressively measurable process $(b_r,r\in[0,T])$ by
\be
b_r:=\frac{\mbb{E}_{\calf_r}\bigl[f^2(r,(Y^1_v)_{v\in[r,T]},Y_r^1,Z_r^1,\psi_r^1)-
f^2(r,(Y^1_v)_{v\in[r,T]},Y_r^1,Z_r^2,\psi_r^1)\bigr]}{|\del Z_r|^2}\bold{1}_{\del Z_r\neq 0}\del Z_r^\top~.\nn
\ee
Since $|b_r|\leq K_M\bigl(1+|Z_r^1|+|Z_r^2|+2||\psi^1_r||^2_{\mbb{L}^2(\nu)}\bigr)$, there exists some constant $C$ such that
$||b||_{\mbb{H}^2_{BMO}}\leq C$ with $C=C(K_\cdot, A)$. Thus one can define an equivalent probability measure $\mbb{Q}$ by
$\frac{d\mbb{Q}}{d\mbb{P}}=\cale_T\bigl(\int_0^\cdot b_r^\top dW_r\bigr)$
where $\cale(\cdot)$ is Dol\'eans-Dade exponential. We have $W^{\mbb{Q}}=W-\int_0^\cdot b_r dr$
and the Poisson measure is unchanged, $\wt{\mu}^{\mbb{Q}}=\wt{\mu}$.
We also have $\frac{d\mbb{P}}{d\mbb{Q}}=\cale_T\bigl(-\int_0^\cdot b_r^\top dW_r^\mbb{Q}\bigr)$.
From Remark~\ref{remark-BMO}, there exists some constant $r^*\in(1,\infty)$ such that
the reverse H\"older inequality holds for both of the $\cale_\cdot(\int_0^\cdot b_r^\top dW_r)$
and $\cale_\cdot(-\int_0^\cdot b_r^\top dW_r^\mbb{Q})$ with power $\bar{r}\in(1,r^*]$.
Define $q_*>1$ by $q_*:=r^*/(r^*-1)$. Note that $(r^*,q_*)$ are solely controlled by $(K_\cdot, A)$.

Under the measure $\mbb{Q}$, we have
\bea
&&\del Y_t=\del \xi+\int_t^T \mbb{E}_{\calf_r}\Bigl[\del f(r)+f^2\bigl(r,(Y^1_v)_{v\in[r,T]},Y^1_r,Z_r^2,\psi_r^1\bigr)
-f^2\bigl(r,(Y^2_v)_{v\in[r,T]},Y^2_r,Z^2_r,\psi^2_r\bigr)\Bigr]dr\nn \\
&&\qquad -\int_t^T \del Z_r dW_r^{\mbb{Q}}-\int_t^T\int_E \del \psi_r(e)\wt{\mu}^{\mbb{Q}}(dr,de),~t\in[0,T].\nn
\eea
%%%%%%%%%%%%%%%%%%%%%%%%%%%%%
{\bf{~[Stability for Y]}}
%%%%%%%%%%%%%%%%%%%%%%%%%%%%%
Applying Ito formula to $\del Y^2$, one obtains
\bea
&&|\del Y_t|^2+\int_t^T |\del Z_r|^2 dr+\int_t^T \int_E |\del \psi_r(e)|^2\mu(dr,de)\nn \\
&&=|\del \xi|^2+\int_t^T 2\del Y_r\mbb{E}_{\calf_r}\Bigl[\del f(r)+f^2\bigl(r,(Y^1_v)_{v\in[r,T]},Y^1_r,Z_r^2,\psi_r^1\bigr)
-f^2\bigl(r,(Y^2_v)_{v\in[r,T]},Y^2_r,Z^2_r,\psi^2_r\bigr)\Bigr]dr\nn \\
&&\quad-\int_t^T 2\del Y_r \del Z_r dW_r^{\mbb{Q}}-\int_t^T \int_E 2\del Y_{r-}\del \psi_r(e)\wt{\mu}^{\mbb{Q}}(dr,de)~.
\label{eq-dely-sq}
\eea 
The last two terms are true $\mbb{Q}$-martingales, which can be checked by reverse H\"older and energy inequalities.
By taking conditional expectation $\mbb{E}^{\mbb{Q}}_{\calf_t}[\cdot]$, one obtains with any $\lambda>0$
\bea
&&|\del Y_t|^2+\mbb{E}^{\mbb{Q}}_{\calf_t}\int_t^T |Z_r|^2 dr+\mbb{E}^{\mbb{Q}}_{\calf_t}\int_t^T ||\del \psi_r||^2_{\mbb{L}^2(\nu)}dr \leq C\mbb{E}_{\calf_t}^{\mbb{Q}}\int_t^T \mbb{E}_{\calf_r}\bigl[||\del Y||_{[r,T]}^2\bigr]dr \nn \\
&&\qquad +\mbb{E}_{\calf_t}^{\mbb{Q}}\Bigl[|\del\xi|^2+\frac{1}{\lambda}\Bigl(\int_t^T \mbb{E}_{\calf_r}|\del f(r)|dr\Bigr)^2
+\lambda ||\del Y||^2_{[t,T]}\Bigr]+\frac{1}{2}\mbb{E}_{\calf_t}^{\mbb{Q}}\int_t^T ||\del \psi_r||^2_{\mbb{L}^2(\nu)}dr\nn
\eea
with some positive constant $C=C(K_\cdot,A)$. Here we have used the fact that $|\del Y_r|\leq \mbb{E}_{\calf_r}
\bigl[||\del Y||_{[r,T]}\bigr]$. Therefore, in particular, 
\bea
&&|\del Y_t|^2\leq \mbb{E}^{\mbb{Q}}_{\calf_t}\Bigl[|\del\xi|^2+\frac{1}{\lambda}\Bigl(\int_t^T \mbb{E}_{\calf_r}|\del f(r)|dr\Bigr)^2
+\lambda||\del Y||^2_{[t,T]}+C\int_t^T \mbb{E}_{\calf_r}\bigl[||\del Y||^2_{[r,T]}\bigr]dr\Bigr]\nn\\
&&= 
\frac{1}{\cale_t}\mbb{E}_{\calf_t}\left[\cale_T
\Bigl(|\del\xi|^2+\frac{1}{\lambda}\Bigl(\int_t^T \mbb{E}_{\calf_r}|\del f(r)|dr\Bigr)^2
+\lambda||\del Y||^2_{[t,T]}+C\int_t^T \mbb{E}_{\calf_r}\bigl[||\del Y||^2_{[r,T]}\bigr]dr\Bigr)\right]\nn
\eea
where $\cale_s:=\cale_s(\int_0^\cdot b_r^\top dW_r)$.
Choosing $\bar{q}\in[q_*,\infty)$, the reverse H\"older inequality yields
\bea
&&|\del Y_t|^{2\bq}\leq C\mbb{E}_{\calf_t}\Bigl[|\del\xi|^{2\bar{q}}+\frac{1}{\lambda^{\bq}}\Bigl(\int_t^T
\mbb{E}_{\calf_r}|\del f(r)|dr\Bigr)^{2\bq}+\Bigl(\int_t^T \mbb{E}_{\calf_r}\bigl[||\del Y||^2_{[r,T]}\bigr]dr\Bigr)^{\bq}
+\lambda^{\bq}||\del Y||^{2\bq}_{[t,T]}
\Bigr]\nn \\
&&\leq C\mbb{E}_{\calf_t}\Bigl[|\del\xi|^{2\bar{q}}+\frac{1}{\lambda^{\bq}}\Bigl(\int_t^T
\mbb{E}_{\calf_r}|\del f(r)|dr\Bigr)^{2\bq}+\int_t^T  ||\del Y||^{2\bq}_{[r,T]}dr
+\lambda^{\bq}||\del Y||^{2\bq}_{[t,T]}
\Bigr]\nn 
\eea
with some $C=C(\bq,K_\cdot, A)$, where in the 2nd line Jensen's inequality was used.
For any $p>2\bq$, applying Doob's maximal inequality, one obtains
\bea
\mbb{E}\Bigl[||\del Y||^p_{[s,T]}\Bigr]
\leq C\mbb{E}\left[|\del \xi|^{p}+\frac{1}{\lambda^\frac{p}{2}}\Bigl(\int_s^T \mbb{E}_{\calf_r}|\del f(r)|dr\Bigr)^p\right]
+C\int_s^T \mbb{E}\Bigl[||\del Y||^p_{[r,T]}\Bigr]dr+C\lambda^\frac{p}{2}\mbb{E}\Bigl[||\del Y||^{p}_{[s,T]}\Bigr]\nn
\eea
with $C=C(p,\bq,K_\cdot, A)$. Choosing $\lambda>0$ small enough so that $C\lambda^\frac{p}{2}<1$, the backward Gronwall inequality implies
\bea
\mbb{E}\Bigl[\sup_{t\in[s,T]}|\del Y_t|^p\Bigr]\leq C\mbb{E}\Bigl[|\del\xi|^p+\Bigl(\int_s^T \mbb{E}_{\calf_r}|\del f(r)|dr\Bigr)^p
\Bigr],~~\forall s\in[0,T]\nn~.
\eea
One sees the last inequality holds for any $p>2q_*$. This proves (\ref{eq-y-stability}). Since $1<q_*\leq \bq$, it also follows that 
\bea
\mbb{E}\Bigl[\sup_{t\in[0,T]}|\del Y_t|^p\Bigr]^\frac{1}{p}\leq \mbb{E}\Bigl[\sup_{t\in[0,T]}|\del Y_t|^{p\bq^2}\Bigr]^\frac{1}{p\bq^2}
\leq C\mbb{E}\Bigl[|\del\xi|^{p\bq^2}+\Bigl(\int_0^T \mbb{E}_{\calf_r}|\del f(r)|dr\Bigr)^{p\bq^2}
\Bigr]^\frac{1}{p\bq^2}
\label{eq-Y-stability}
\eea
with $C=C(p,\bq,K_\cdot,A)$ for any $p\geq 2$.
\\
\\
{\bf{[Stability for $Z$ and $\psi$]}}
From (\ref{eq-dely-sq}), one has with $C=C(K_\cdot,A)$, 
\bea
&&|\del Y_t|^2+\int_t^T |\del Z_r|^2dr+\int_t^T \int_E |\del \psi_r(e)|^2\mu(dr,de)\nn \\
&&\leq |\del \xi|^2+\Bigl(\int_t^T \mbb{E}_{\calf_r}|\del f(r)|dr\Bigr)^2+
||\del Y||^2_{[t,T]}+C\int_t^T\mbb{E}_{\calf_r}\bigl[||\del Y||^2_{[r,T]}\bigr]dr\nn \\
&&\quad+C\int_t^T |\del Y_r|||\del\psi_r||_{\mbb{L}^2(\nu)}dr-\int_t^T 2\del Y_r\del Z_r dW_r^{\mbb{Q}}
-\int_t^T \int_E 2\del Y_{r-}\del\psi_r(e)\wt{\mu}^{\mbb{Q}}(dr,de)\nn~.
\eea
For any $p\geq 2$, applying Burkholder-Davis-Gundy inequality\footnote{See, for example, Theorem 48 in IV.4. of \cite{Protter}.}
and Lemma~\ref{lemma-Shiryayev}, 
one can show that there exists some constant $C=C(p,K_\cdot, A)$ such that 
\bea
&&\mbb{E}^{\mbb{Q}}\Bigl[\Bigl(\int_0^T |\del Z_r|^2 dr\Bigr)^\frac{p}{2}\Bigr]
+\mbb{E}^{\mbb{Q}}\Bigl[\Bigl(\int_0^T\int_E |\del\psi_r(e)|^2\mu(dr,de)\Bigr)^\frac{p}{2}\Bigr]\nn \\
&&\leq C\mbb{E}^{\mbb{Q}}\Bigl[|\del\xi|^p+\Bigl(\int_0^T \mbb{E}_{\calf_r}|\del f(r)|dr\Bigr)^p+\sup_{r\in[0,T]}\mbb{E}_{\calf_r}\bigl[||\del Y||^p_{[0,T]}\bigr]+||\del Y||^p_{[0,T]}\Bigr]~.\nn
\eea
Taking $\bq\geq q_*$, the reverse H\"older and Doob's maximal inequalities give
\bea
&&\mbb{E}^{\mbb{Q}}\Bigl[\Bigl(\int_0^T |\del Z_r|^2 dr\Bigr)^\frac{p}{2}\Bigr]^\frac{1}{p}
+\mbb{E}^{\mbb{Q}}\Bigl[\Bigl(\int_0^T\int_E |\del\psi_r(e)|^2\mu(dr,de)\Bigr)^\frac{p}{2}\Bigr]^\frac{1}{p}\nn \\
&&\leq C\mbb{E}\Bigl[|\del\xi|^{p\bq}+\Bigl(\int_0^T \mbb{E}_{\calf_r}|\del f(r)|dr\Bigr)^{p\bq}+\sup_{r\in[0,T]}\mbb{E}_{\calf_r}\bigl[||\del Y||_{[0,T]}^p\bigr]^\bq+||\del Y||^{p\bq}_{[0,T]}\Bigr]^\frac{1}{p\bq}\nn \\
&&\leq C\mbb{E}\Bigl[|\del\xi|^{p\bq}+\Bigl(\int_0^T \mbb{E}_{\calf_r}|\del f(r)|dr\Bigr)^{p\bq}
+||\del Y||^{p\bq}_{[0,T]}\Bigr]^\frac{1}{p\bq}.\nn
\eea
The reverse H\"older inequality implies $||Z||_{\mbb{H}^p}+||\psi||_{\mbb{J}^P}\leq C\bigl(||Z||_{\mbb{H}^{p\bq}(\mbb{Q})}+
||\psi||_{\mbb{J}^{p\bq}(\mbb{Q})}\bigr)$. Thus the estimate of (\ref{eq-Y-stability}) and Lemma~\ref{lemma-Shiryayev} give
\bea
%\mbb{E}\Bigl[\Bigl(\int_0^T |\del Z_r|^2 dr\Bigr)^\frac{p}{2}\Bigr]^\frac{1}{p}+
%\mbb{E}\Bigl[\Bigl(\int_0^T\int_E |\del \psi_r(e)|^2\mu(dr,de)\Bigr)^\frac{p}{2}\Bigr]^\frac{1}{p} 
||\del Y||_{\mbb{S}^p}+||\del Z||_{\mbb{H}^p}+||\del \psi||_{\mbb{J}^p}
\leq C\mbb{E}\Bigl[|\del \xi|^{p\bq^2}+\Bigl(\int_0^T \mbb{E}_{\calf_r}|\del f(r)|dr \Bigr)^{p\bq^2}\Bigr]^\frac{1}{p\bq^2}\nn
\eea
for any $p\geq 2$ and $\bq\geq q_*$ with some positive constant $C=C(p,\bq,K_\cdot, A)$. 
\end{proof}
\end{proposition}

We also have the following relation.
\begin{lemma}
\label{lemma-sinfty-uniqueness}
Under the same conditions used in Proposition~\ref{prop-stability}, one has
\bea
||\del Z||_{\mbb{H}^2_{BMO}}+||\del \psi||_{\mbb{J}^2_{BMO}}\leq
C\Bigl(||\del Y||_{\mbb{S}^\infty}+||\del\xi||_{\infty}+\sup_{t\in\calt^T_0}\Bigl|\Bigl|
\mbb{E}_{\calf_t}\int_t^T |\del f(r)|dr\Bigr|\Bigr|_{\infty}\Bigr) \nn
\eea
with some positive constant $C=C(K_\cdot,A)$.
\begin{proof}
It follows from a simple modification of Lemma~3.3 (a) of \cite{FT-Qexp}.
\end{proof}
\end{lemma}

Combining the results in this section, we obtain the uniqueness.
\begin{corollary}
\label{corollary-uniqueness-nonM}
Under Assumptions~\ref{assumption-structure-nonM} and \ref{assumption-LLC-nonM}, 
if the ABSDE (\ref{eq-qg-bsde-nonM}) has a bounded solution $(Y,Z,\psi)\in \mbb{S}^\infty\times \mbb{H}^2\times \mbb{J}^2$,
then it is unique with respect to the norm $\mbb{S}^\infty\times \mbb{H}^2_{BMO}\times \mbb{J}^2_{BMO}$.
\begin{proof}
Proposition~\ref{prop-stability} implies the uniqueness of $Y$ in $\mbb{S}^p,~\forall p\geq 2$, in particular.
This also implies the uniqueness with respect to $\mbb{S}^\infty$. If not, there exists some $c>0$
such that $||\del Y||_{\mbb{S}^\infty}=c$, which implies for any $0<b<c$, there exists
a strictly positive constant $a>0$ such that
$\mbb{P}(\sup_{t\in[0,T]}|\del Y_t|>b)=a$. This yields $||\del Y||_{\mbb{S}^p}^p>b^p a>0$, which is a contradiction.
Thus the assertion follows from Proposition~\ref{prop-stability} and Lemma~\ref{lemma-sinfty-uniqueness}.
\end{proof}
\end{corollary}

\begin{remark}
For quadratic BSDEs, allowing the anticipated components of $(Z,\psi)$ in the driver $f$ seems very hard.
In fact, we cannot derive the stability result similar to Proposition~\ref{prop-stability}.
This is because that the use of the reverse H\"older inequality makes 
the power of $(|Z|,|\psi|)$ different in the left and right hand sides in the relevant inequalities
after aligning the probability measure of the conditional expectations to a single one. 
The anticipated component for $Y$ is an exceptional case, where we can 
remove one conditional expectation by the simple fact $Y_t=\mbb{E}^{\mbb{Q}}[Y_t|\calf_t]$.
Note that the Proposition~\ref{prop-stability} is necessary also for the non-Markovian settings in 
Section~\ref{sec-non-Markovian}. 
In the absence of the stability result, the convergence using the monotone sequence would be the last hope.
However, to the best of our knowledge, no comparison principle is known
in the presence of anticipated components of the control variables $(Z,\psi)$.
\end{remark}

%%%%%%%%%%%%%%%%%%%%%%%%%%%%%%%%%%%%%%%%%%%%%%%%
\section{Existence in a Markovian Setup}
%%%%%%%%%%%%%%%%%%%%%%%%%%%%%%%%%%%%%%%%%%%%%%%%
Let us now provide the existence result for a Markovian setting.
We introduce the following forward process, for $s\in[0,T]$, 
\bea
X_s^{t,x}=x+\int_t^{s\vee t} b(r,X_r^{t,x})dr+\int_t^{s\vee t} \sigma(r,X_r^{t,x})dW_r
+\int_t^{s\vee t}\int_E\gamma(r,X_{r-}^{t,x},e)\wt{\mu}(dr,de)
\label{eq-x}
\eea
where $x\in \mbb{R}^n$ and $b:[0,T]\times \mbb{R}^n \rightarrow \mbb{R}^n$, 
$\sigma:[0,T]\times \mbb{R}^n\rightarrow \mbb{R}^{n\times d}$,
$\gamma:[0,T]\times \mbb{R}^n\times E\rightarrow \mbb{R}^{n\times k}$ are non-random measurable functions.
Note that $X_s^{t,x}\equiv x$ for $s\leq t$.
\begin{assumption}
\label{assumption-X} There exists a positive constant $K$ such that\\
(i) $|b(t,0)|+|\sigma(t,0)|\leq K$ uniformly in $t\in[0,T]$. \\
(ii) $\sum_{i=1}^k |\gamma^i(t,0,e)|\leq K(1\wedge |e|)$ uniformly in $(t,e)\in[0,T]\times \mbb{R}_0$.\\
(iii) uniformly in $t\in[0,T], x,x^\prime\in \mbb{R}^n,e\in \mbb{R}_0$,
\bea
&&|b(t,x)-b(t,x^\prime)|+|\sigma(t,x)-\sigma(t,x^\prime)|\leq K|x-x^\prime|,\nn \\
&&\sum_{i=1}^k|\gamma^i(t,x,e)-\gamma^i(t,x^\p,e)|\leq K(1\wedge|e|)|x-x^\p|~.\nn
\eea
\end{assumption}

\begin{lemma}
\label{lemma-x-continuity}
Under Assumption~\ref{assumption-X}, there exists a unique solution to (\ref{eq-x}) for each $(t,x)$
which satisfies
for any $(t,x),(t,x^\p)\in[0,T]\times \mbb{R}^n$ and $p\geq 2$,
\bea
&(a)& \mbb{E}\Bigl[\sup_{s\in[0,T]}|X_s^{t,x}|^p\Bigr]\leq C(1+|x|^p) \nn \\
&(b)& \mbb{E}\Bigl[\sup_{s\leq u\leq (s+h)\wedge T}|X_s^{t,x}-X_u^{t,x}|^p\Bigr]\leq C(1+|x|^p)h,  ~\forall s\in[0,T] \nn \\
&(c)& \mbb{E}\Bigl[\sup_{s\in[0,T]}|X_s^{t,x}-X_s^{t^\p,x^\p}|^p\Bigr]\leq C\Bigl(|x-x^\p|^p+
(1+[|x|\vee|x^\p|]^p)|t-t^\p|\Bigr)\nn 
\eea
with some constant $C=C(p,K,T)$.
\begin{proof}
They are the standard estimates for the Lipschitz SDEs. See, for example, Theorem 4.1.1~\cite{Delong}.
For the selfcontainedness, we give a proof in Appendix~\ref{ap-x-continuity} for regularities.
\end{proof}
\end{lemma}

We are interested in the Markovian anticipated BSDE associated with $(X^{t,x}_v)_{v\in[0,T]}$:
\bea
&&Y_s^{t,x}=\xi(X_T^{t,x})+\int_s^T \bold{1}_{r\geq t}\mbb{E}_{\calf_r}f\Bigl(r,X_r^{t,x},(Y^{t,x}_v)_{v\in[r,T]},
Y_r^{t,x},Z_r^{t,x},\psi_r^{t,x}\Bigr)dr\nn \\
&&\qquad -\int_s^T Z_r^{t,x}dW_r-\int_s^T \int_E \psi_r^{t,x}(e)\wt{\mu}(dr,de)~,
\label{eq-qg-bsde-M}
\eea
where $f:[0,T]\times\mbb{R}^n\times  \mbb{D}[0,T]\times \mbb{R}\times \mbb{R}^{1\times d}\times \mbb{L}^2(E,\nu)
\rightarrow \mbb{R}$ and $\xi:\mbb{R}^n\rightarrow \mbb{R}$ are non-random measurable functions.
Note that $(Y^{t,x}_s,Z^{t,x}_s,\psi^{t,x}_s)\equiv (Y^{t,x}_t,0,0)$ for $s\leq t$.

\begin{assumption}
\label{assumption-structure-M}
(i)The driver $f$ is a map such that for every $(x,y,z,\psi)\in \mbb{R}^n\times \mbb{R}\times \mbb{R}^{1\times d}\times 
\mbb{L}^2(E,\nu)$ and any \cadlag $\mbb{F}$-adapted process $(Y_v)_{v\in[0,T]}$, the process
$\bigl( \mbb{E}_{\calf_t}f(t,x,(Y_v)_{v\in[t,T]}, y,z,\psi), t\in[0,T]\bigr)$ is $\mbb{F}$-progressively measurable. \\
(ii)For every $(x,q,y,z,\psi)\in \mbb{R}^n\times \mbb{D}[0,T]\times \mbb{R}\times \mbb{R}^{1\times d}\times \mbb{L}^2(E,\nu)$,
there exist constants $\beta, \del\geq 0$, $\gamma>0$ and a positive non-random function $l:[0,T]\rightarrow \mbb{R}$ such that
\bea
&&-l_t-\del\bigl(\sup_{v\in[t,T]}|q_v|\bigr)-\beta|y|-\frac{\gamma}{2}|z|^2-\int_E j_\gamma (-\psi(e))\nu(de) ~\leq~ f\bigl(t,x,(q_v)_{v\in[t,T]},y,z,\psi\bigr)\nn \\
&&\qquad \leq l_t+\del\bigl(\sup_{v\in[t,T]}|q_v|\bigr)+\beta|y|+\frac{\gamma}{2}|z|^2
+\int_E j_\gamma (\psi(e))\nu(de) \nn
\eea
$dt$-a.e. $t\in[0,T]$, where $j_\gamma(u)=\frac{1}{\gamma}(e^{\gamma u}-1-\gamma u)$.\\
(iii) $||\xi(\cdot)||_{\infty},~\sup_{t\in[0,T]}(l_t)<\infty$.
\end{assumption}

\begin{assumption}
\label{assumption-LLC-M}
For each $M>0$, and for every $(x,q,y,z,\psi), (x^\p,q^\p,y^\p,z^\p,\psi^\p)\in \mbb{R}^n\times \mbb{D}[0,T]
\times \mbb{R}\times \mbb{R}^{1\times d}\times \mbb{L}^2(E,\nu)$ satisfying
$|y|,|y^\p|,||\psi||_{\mbb{L}^\infty(\nu)},||\psi^\p||_{\mbb{L}^\infty(\nu)}$, $\sup_{v\in[0,T]}|q_v|,
\sup_{v\in[0,T]}|q^\p_v|\leq M$, there exist some positive constants $K_M$ (depending on $M$) 
and $K_\xi\geq 0, \rho\geq 0,~\alpha\in(0,1]$  such that, for $dt$-a.e. $t\in[0,T]$,
\bea
&& \bullet~ |f(t,x,(q_v)_{v\in[t,T]},y,z,\psi)-f(t,x,(q^\p_v)_{v\in[t,T]},y^\p,z^\p,\psi^\p)|\nn \\
&&\qquad\qquad \leq K_M\Bigl(\sup_{v\in[t,T]}|q_v-q^\p_v|+|y-y^\p|+||\psi-\psi^\p||_{\mbb{L}^2(\nu)}\Bigr)\nn \\
&&\qquad\qquad +K_M\bigl(1+|z|+|z^\p|+||\psi||_{\mbb{L}^2(\nu)}+||\psi^\p||_{\mbb{L}^2(\nu)}\bigr)|z-z^\p|~,\nn \\
&& \bullet~ |f(t,x,(q_v)_{v\in[t,T]},y,z,\psi)-f(t,x^\p,(q_v)_{v\in[t,T]},y,z,\psi)|\nn \\
&&\qquad\qquad \leq K_M\bigl(1+[|x|\vee|x^\p|]^\rho+|z|^2+||\psi||^2_{\mbb{L}^2(\nu)}\bigr)|x-x^\p|^{\alpha}~,\nn
\eea
and $|\xi(x)-\xi(x^\p)|\leq K_\xi|x-x^\p|^{\alpha}$.
\end{assumption}

\begin{proposition}
\label{prop-u-continuity}
Under Assumptions~\ref{assumption-X}, \ref{assumption-structure-M} and \ref{assumption-LLC-M}, 
suppose that there exists a bounded solution $(Y^{t,x},Z^{t,x},\psi^{t,x})\in \mbb{S}^\infty\times \mbb{H}^2\times \mbb{J}^2$
for each $(t,x)\in[0,T]\times \mbb{R}^n$. Then the solution is unique and 
$(Y^{t,x},Z^{t,x},\psi^{t,x})\in \mbb{S}^\infty\times \mbb{H}^2_{BMO}\times \mbb{J}^2_{BMO}$
with the norm solely controlled by $A=(||\xi||_{\infty},sup_{t\in[0,T]}l_t, \del,\beta,\gamma,T)$,
which is, in particular, independent of $(t,x)\in[0,T]\times \mbb{R}^n$.

Moreover, if $Y_t^{t,x}$ is a deterministic map in $(t,x)$, 
the map $u:[0,T]\times \mbb{R}^n\rightarrow \mbb{R}$ defined by $u(t,x):=Y^{t,x}_t$
satisfies for any pair of $(t,x),(\tp,\xp)\in[0,T]\times \mbb{R}^n$,
\bea
|u(t,x)-u(t^\p,x^\p)|\leq C\Bigl(1+[|x|\vee|x^\p|]^\rho\Bigr)\Bigl(|x-x^\p|^\alpha+
\bigl(1+[|x|\vee|x^\p|]^\alpha\bigr)|t-\tp|^\frac{1}{2p\bq^2}\Bigr)\nn
\eea
with some constant $C=C(\alpha,\rho,p,\bq,K_\xi,K,K_\cdot, A)$ 
for any $p\geq 2$ and $\bq\in[q_*,\infty)$ such that $\alpha p\bq^2\geq 1$, where $q_*>1$ is some constant determined by $(K_\cdot,A)$.
\begin{proof}
The first part follows from Lemmas~\ref{lemma-z-psi-bound}, \ref{lemma-y-bound} 
and Corollary~\ref{corollary-uniqueness-nonM}.

Let us assume $\tp\leq t$ without loss of any generality.
Put $\del Y:=Y^{t,x}-Y^{\tp,\xp}$,
\bea
\del f(r)&:=&\bold{1}_{r\geq t}f(r,X_r^{t,x},(Y^{t,x}_v)_{v\in[r,T]},\Theta_r^{t,x})
-\bold{1}_{r\geq \tp} f(r,X_r^{\tp,\xp},(Y^{t,x}_v)_{v\in[r,T]},\Theta_r^{t,x})\nn \\
&=&\bold{1}_{r\geq t}\Bigl(f(r,X_r^{t,x},(Y^{t,x}_v)_{v\in[r,T]},\Theta_r^{t,x})-f(r,X_r^{\tp,\xp},(Y^{t,x}_v)_{v\in[r,T]},\Theta_r^{t,x})\Bigr)\nn \\
&&-\bold{1}_{\tp\leq r\leq t}f(r,X_r^{\tp,\xp},(Y^{t,x}_v)_{v\in[r,T]},Y_t^{t,x},0,0)~,\nn
\eea
and $\del\xi=\xi(X_T^{t,x})-\xi(X_T^{\tp,\xp})$. By Proposition~\ref{prop-stability}, for any $p\geq 2,\bq\in[q_*,\infty)$,
\bea
|u(t,x)-u(\tp,\xp)|&\leq& \mbb{E}\Bigl[\sup_{s\in[0,T]}|Y^{t,x}_s-Y^{\tp,\xp}_s|^p\Bigr]^\frac{1}{p}\nn \\
&\leq &C\mbb{E}\Bigl[|\del\xi|^{p\bq^2}+\Bigl(\int_0^T \mbb{E}_{\calf_r}|\del f(r)| dr\Bigr)^{p\bq^2}\Bigr]^\frac{1}{p\bq^2} \nn
\eea
with $C=C(p,\bq,K_\cdot,A)$. 
The universal bounds of Lemmas~\ref{lemma-z-psi-bound} and \ref{lemma-y-bound} imply that
$||Y^{t,x}||_{\mbb{S}^\infty}$, $||Z^{t,x}||_{\mbb{H}^2_{BMO}},~||\psi^{t,x}||_{\mbb{J}^2_{BMO}}\leq C$
with some $C=C(A)$ uniformly in $(t,x)$. Thus one can apply fixed $K_M$ for the whole range
in Assumption~\ref{assumption-LLC-M} provided $M$ is chosen large enough.
It follows that
\bea
\mbb{E}_{\calf_r}|\del f(r)|&\leq & \bold{1}_{r\geq t} K_M\Bigl(1+\bigl[|X_r^{t,x}|\vee |X_r^{\tp,\xp}|\bigr]^\rho
+|Z_r^{t,x}|^2+||\psi_r^{t,x}||^2_{\mbb{L}^2(\nu)}\Bigr)|X_r^{t,x}-X_r^{\tp,\xp}|^\alpha \nn \\
&+& \bold{1}_{\tp\leq r\leq t}\bigl(l+\del \mbb{E}_{\calf_r}[||Y^{t,x}||_{[r,T]}]+\beta |Y_t^{t,x}|\bigr)~.\nn
\eea
Hence,  using the boundedness of $Y^{t,x}$ and Cauchy-Schwartz inequality,  one obtains
\bea
&&\mbb{E}\Bigl[\Bigl(\int_0^T \mbb{E}_{\calf_r}|\del f(r)|dr\Bigr)^{p\bq^2}\Bigr]^\frac{1}{p\bq^2}\nn \\
&&\leq C \mbb{E}\left[1+\bigl[||X^{t,x}||_{[0,T]}\vee ||X^{\tp,\xp}||_{[0,T]}\bigr]^{2\rho p\bq^2}+\Bigl(\int_0^T |Z_r^{t,x}|^2+||\psi_r^{t,x}||^2_{\mbb{L}^2(\nu)}dr\Bigr)^{2p\bq^2}\right]^\frac{1}{2p\bq^2}\nn \\
&&\quad \times \mbb{E}\Bigl[||X^{t,x}-X^{\tp,\xp}||_{[0,T]}^{2\alpha p\bq^2}\Bigr]^\frac{1}{2p\bq^2}+C|t-\tp|~.\nn
\eea
Note here that, by the energy inequality~\footnote{See, for example, Lemma 2.2~\cite{FT-Qexp}. As for a simple proof, see Lemma 9.6.5~\cite{Cvitanic}.}, 
the following relation holds:
\bea
\mbb{E}\Bigl[\Bigl(\int_0^T |Z_r^{t,x}|^2+||\psi_r^{t,x}||^2_{\mbb{L}^2(\nu)}dr\Bigr)^{2p\bq^2}\Bigr]^\frac{1}{2p\bq^2}\leq C
\eea
where the constant $C$ depends only on $(||Z||_{\mbb{H}^2_{BMO}}, ||\psi||_{\mbb{J}^2_{BMO}})$ and $p\bq^2$.
Using Lemma~\ref{lemma-x-continuity}(a) and (c), one obtains the desired regularity.
The contribution from $\del\xi$ can be computed similarly.
\end{proof}
\end{proposition}
\begin{remark}
Under the conditions of the above proposition, we have, for each $s\in[0,T]$, $Y_s^{t,x}=Y_s^{s,X_s^{t,x}}=u(s,X_s^{t,x})$ a.s. 
due to the uniqueness of solution $Y^{t,x}$. Furthermore, since the function $u$ is jointly continuous, 
$u(s,X_s^{t,x})_{s\in[0,T]}$ is \cadlag $\mbb{F}$-adapted. Thus, Chapter 1, Theorem 2  of  \cite{Protter}
implies that $Y^{t,x}_s=u(s,X_s^{t,x})$  $\forall s\in[0,T]$ a.s.
\end{remark}

We now introduce a sequence of regularized anticipated BSDEs with $m\in \mbb{N}$:
\bea
Y^{m,t,x}_s&=&\xi(X_T^{t,x})+\int_s^T \bold{1}_{r\geq t}\mbb{E}_{\calf_r}f_m\bigr(r,X_r^{t,x},(Y^{m,t,x}_v)_{v\in[r,T]}, 
Y_r^{m,t,x},Z_r^{m,t,x},\psi_r^{m,t,x}\bigr)dr\nn \\
&&-\int_s^T Z_r^{m,t,x}dW_r-\int_s^T\int_E \psi_r^{m,t,x}(e)\wt{\mu}(dr,de)
\label{eq-absde-regularized}
\eea
where $f_m$ is defined by, $\forall (r,x, q,y,z,\psi)\in[0,T]\times \mbb{R}^n \times \mbb{D}[0,T]\times\mbb{R}\times \mbb{R}^{1\times d}
\times \mbb{L}^2(E,\nu)$,
\bea
f_m\bigl(r,x, (q_s)_{v\in[r,T]},y,z,\psi\bigr):=f\bigl(r,x,(\varphi_m(q_s))_{v\in[r,T]},\varphi_m(y),\varphi_m(z),\varphi_m(\psi\circ\zeta_m)\bigr)~.
\label{eq-fm-def}
\eea
Here, we have used a simple truncation function 
\bea
\varphi_m(x):=\begin{cases}
-m \quad{\text{for $x\leq -m$}}\\
~x  \qquad{\text{for $|x|\leq m$}}\\
~m \quad~~{\text{for $x\geq m$}}
\end{cases} \nn
\eea
and a cutoff function $\psi\circ \zeta_m(e):=\psi(e)\bold{1}_{|e|\geq 1/m}$, which are applied component-wise for $z,\psi$.

\begin{lemma}
\label{lemma-fm-regularized}
Suppose that the driver $f$ satisfies Assumptions~\ref{assumption-structure-M} and \ref{assumption-LLC-M}.
Then, $(f_m)_{m\in \mbb{N}}$ also satisfy Assumptions~\ref{assumption-structure-M} and \ref{assumption-LLC-M}
uniformly in $m\in \mbb{N}$. Moreover, for each $m\in\mbb{N}$, the driver $f_m$ is a.e. bounded 
and globally Lipschitz continuous with respect to $(q,y,z,\psi)$ in the sense of Assumption~\ref{assumption-Lipschitz}.
\begin{proof}
With $|\varphi_m(x)|\leq |x|$, $|\varphi_m(x)-\varphi_m(\xp)|\leq |x-\xp|$ and use the convexity of the function $j_\gamma(\cdot)$,
the first claim is obvious.
By denoting $C_m:=\max_{1\leq k\leq 1}\int_{|e|\geq 1/m}\nu^i(de)<\infty$, 
one sees $|f_m|\leq \sup_{t\in[0,T]}l_t+(\del+\beta)m+\frac{\gamma}{2}dm^2+k j_\gamma(m)C_m$ a.e. by the structure condition.
By noticing the fact that
\bea
||\varphi_m(\psi\circ\zeta_m)||_{\mbb{L}^2(\nu)}^2\leq \sum_{i=1}^k m^2\int_{|e|\geq 1/m}\nu^i(de) \leq k m^2C_m\nn
\eea
the global Lipschitz continuity can be confirmed easily.
\end{proof}
\end{lemma}

\begin{lemma}
\label{lemma-Ym-deterministic}
There exists a unique solution $(Y^{m,t,x}, Z^{m,t,x}, \psi^{m,t,x})$ to (\ref{eq-absde-regularized})
satisfying 
\be
||Y^{m,t,x}||_{\mbb{S}^\infty}, ~||Z^{m,t,x}||_{\mbb{H}^2_{BMO}}, ~||\psi^{m,t,x}||_{\mbb{J}^2_{BMO}}\leq C\nn
\ee
with some constant $C=C(A)$, depending only on those relevant for the universal bound,
uniformly in $(m,t,x)\in\mbb{N}\times [0,T]\times \mbb{R}^n$.
Moreover,  $(Y^{m,t,x}_s, Z^{m,t,x}_s, \psi^{m,t,x}_s; s\in[t,T])$ is adapted to the 
$\sigma$-algebra  $\calf_s^t$ generated by $(W,\mu)$ after $t$, that is,
$\calf_s^t=\sigma\bigl(W_u-W_t, ~\mu((t,u],\cdot); ~t\leq u\leq s \bigr)$ for each $s\in[t,T]$. In particular,
$Y^{m,t,x}_t$ is deterministic in $(t,x)$.
\begin{proof}
Thanks to Lemma~\ref{lemma-fm-regularized}, Proposition~\ref{prop-Lipschitz} is applicable to (\ref{eq-absde-regularized}),
which implies  that there exists a unique solution $(Y^{m,t,x},Z^{m,t,x},\psi^{m,t,x})\in\calk^2[0,T]$
of (\ref{eq-absde-regularized}). Since $|\xi|$ and $|f_m|$ are bounded, we actually have $Y^{m,t,x}\in \mbb{S}^\infty$.
Therefore, Lemmas~\ref{lemma-fm-regularized}, \ref{lemma-z-psi-bound}
and \ref{lemma-y-bound} imply the desired bound
\be
||Y^{m,t,x}||_{\mbb{S}^\infty}, ||Z^{m,t,x}||_{\mbb{H}^2_{BMO}}, ||\psi^{m,t,x}||_{\mbb{J}^2_{BMO}}\leq C\nn
\ee
{\it uniformly} in $(m,t,x)\in\mbb{N}\times [0,T]\times \mbb{R}^n$. This proves the first part.

We can prove the latter claims by following the same idea given in Proposition 4.2~\cite{ElKaroui-Quenez}
or Theorem 9.5.6~\cite{Cvitanic}. Consider the shifted Brownian motion and Poisson random measure $(W^\prime, \mu^\prime)$ defined by
$W^\prime_s:=W_{t+s}-W_t,~\mu^\prime((0,s],\cdot):=\mu((t,t+s],\cdot), ~0\leq s\leq T-t$,
as well as their associated filtration $\calf_s^\prime:=\calf^t_{t+s}$.
Let $(X_s^{\prime (0,x)}, 0\leq s\leq T-t)$ be the solution to the following SDE:
\bea
X_s^{\prime (0,x)}=x+\int_0^s b(r+t,X_r^{\prime (0,x)})dr+\int_0^s \sigma(r+t,X_r^{\prime(0,x)})dW_r^\prime
+\int_0^s\int_E \gamma(r+t,X_{r+t-}^{\prime(0,x)},e)\wt{\mu}^\prime(dr,de)~\nn
\eea
where $\wt{\mu}^\prime$ is the compensated measure for $\mu^\prime$.
By the strong uniqueness of the SDE, $X_{s-t}^{\prime(0,x)}=X_s^{t,x}$ for $t\leq s\leq T$ $\mbb{P}$-a.s.
Hence $X_s^{t,x}$ is $\calf_s^t~(=\calf^\prime_{s-t})$-measurable. 

Similarly, let us consider the Lipschitz ABSDE for $s\in[0,T-t]$;
\bea
Y_s^{\prime}&=&\xi(X_{T-t}^{\prime(0,x)})+\int_s^{T-t}\mbb{E}_{\calf_r^\prime}f_m(r+t, X_r^{\prime(0,x)},
(Y_v^\prime)_{v\in[r,T-t]},Y_r^\prime, Z_r^\prime,\psi_r^\prime)dr\nn \\
&&-\int_s^{T-t}Z_r^\prime dW_r^\prime
-\int_s^{T-t}\int_E \psi_r^{\prime}(e)\wt{\mu}^\prime(dr,de)~,\nn
\eea
where $(Y^\prime,Z^\prime,\psi^\prime)$ is the unique solution with respect to the filtration $(\calf_s^\prime)_{s\in[0,T-t]}$
by Proposition~\ref{prop-Lipschitz}. Note here that the conditional expectation $\mbb{E}_{\calf_r^\prime}[\cdot]=\mbb{E}_{\calf_{r+t}^t}[\cdot]$
applied to the driver can be replaced by $\mbb{E}_{\calf_{r+t}}[\cdot]$ since the arguments of $f_m$
are adapted to $(\calf^\prime_s)_{s\in[0,T-t]}$ and hence independent of $\calf_t$.
Changing the integration variable to $r+t\rightarrow r\in[t,T]$, and using the fact that 
$dW_{r-t}^\prime=dW_{r}$ and $\wt{\mu}^\prime (d(r-t),de)=\wt{\mu}(dr,de)$,
one obtains
\bea
Y_{s}^\prime&=&\xi(X_{T-t}^{\prime(0,x)})+\int_{s+t}^T\mbb{E}_{\calf_r}f_m(r,X_{r-t}^{\prime(0,x)},(Y^\prime_{v-t})_{v\in[r,T]},
Y^\prime_{r-t},
Z^\prime_{r-t},\psi^\prime_{r-t})dr\nn \\
&&-\int_{s+t}^T Z_{r-t}^\prime dW_r-\int_{s+t}^T\int_E \psi^{\prime}_{r-t}(e)\wt{\mu}(dr,de)\nn.
\eea
Since $X_{s-t}^{\prime(0,x)}=X_s^{t,x}$, one sees that $(Y^\prime_{s-t}, Z^\prime_{s-t}, \psi^\prime_{s-t}; s\in[t,T])$ is a solution 
to (\ref{eq-absde-regularized}) on $[t,T]$. 
Since the Lipschitz ABSDE has a unique solution by Proposition~\ref{prop-Lipschitz}
(Alternatively, one can  use the stability result in Proposition~\ref{prop-stability}),
$(Y^\prime_{s-t}, Z^{\prime}_{s-t}, \psi^\prime_{s-t})=(Y^{m,t,x}_s, Z^{m,t,x}_s,\psi^{m,t,x}_s)$ a.s.
for every $s\in[t,T]$. Thus,  $Y^{m,t,x}_s$ is $\calf_s^t~(=\calf_{s-t}^\prime)$-measurable.
In particular, $Y_t^{m,t,x}$ is $\calf_t^{t}(=\calf^\prime_0)$-measurable and hence deterministic by Blumenthal's $0$-$1$ law.
\end{proof}
\end{lemma}

We now provide our first main result.
\begin{theorem}
\label{th-existence-M}
Under Assumptions~\ref{assumption-X}, \ref{assumption-structure-M} and \ref{assumption-LLC-M},
there exists a unique solution $(Y^{t,x},Z^{t,x},\psi^{t,x})\in \mbb{S}^\infty\times \mbb{H}^2_{BMO}\times \mbb{J}^2_{BMO}$
to the ABSDE (\ref{eq-qg-bsde-M}) for each $(t,x)\in[0,T]\times \mbb{R}^n$.
\begin{proof}
Since the uniqueness follows from the first part of Proposition~\ref{prop-u-continuity}, it suffices to prove the existence.
Lemmas~\ref{lemma-fm-regularized}, \ref{lemma-Ym-deterministic} and Proposition~\ref{prop-u-continuity}
imply that the deterministic map $u_m:[0,T]\times \mbb{R}^n\rightarrow \mbb{R}$
defined by $u_m(t,x):=Y^{m,t,x}_t$ satisfies the local H\"older continuity 
uniformly in $m$ with $C=C(\alpha,\rho,p,\bq,K_\xi,K,K_\cdot,A)$ such that
\bea
|u_m(t,x)-u_m(\tp,\xp)|\leq C\Bigl(1+[|x|\vee|x^\p|]^\rho\Bigr)\Bigl(|x-x^\p|^\alpha+
\bigl(1+[|x|\vee|x^\p|]^\alpha\bigr)|t-\tp|^\frac{1}{2p\bq^2}\Bigr)~.
\label{eq-um-continuity}
\eea
From Lemma~\ref{lemma-Ym-deterministic}, 
it is also clear that $\sup_{m\geq 1}\sup_{(t,x)\in[0,T]\times \mbb{R}^n}|u_m(t,x)| \leq C$.

Let us now confirm the compactness result for $(u_m)_{m\in\mbb{N}}$.
By defining the compact set $\mbb{K}_j$ with $j\in \mbb{N}$ by 
$\mbb{K}_j:=[0,T]\times \ol{B}_j(\mbb{R}^n)\subset\mbb{R}^{n+1}$, we have
$\bigcup_{j=1}^\infty \mbb{K}_j=[0,T]\times \mbb{R}^n$. Here, $\ol{B}_j(\mbb{R}^n)$ is 
a closed ball in $\mbb{R}^n$ of radius $j$ centered at the origin.
Arzel\`a-Ascoli theorem (see, Section 10.1~\cite{Royden}) tells that
there exists a subsequence $(m^{(1)})\subset (m)$ such that, $\exists u^{(1)}\in C(\mbb{K}_1)$,
$(u_{m^{(1)}})$ converges uniformly to $u^{(1)}$ on $\mbb{K}_1$.
Since the sequence $(u_{m^{(1)}})$ is also bounded and equicontinuous, there exists a further subsequence
$(m^{(2)})\subset (m^{(1)})$ such that, $\exists u^{(2)}\in C(\mbb{K}_2)$,
$(u_{m^{(2)}})$ converges uniformly to $u^{(2)}$ on $\mbb{K}_2$. By construction, it is clear that $u^{(2)}|_{\mbb{K}_1}=u^{(1)}$.
Continue the above procedures and construct a diagonal sequence as
\bea
(m^{(m)})_{m\geq 1}:=\{1^{(1)},2^{(2)},\cdots, j^{(j)},\cdots\}~.\nn
\eea
From Lemma 2 in Section 10.1~\cite{Royden} implies that there exists a subsequence $(m^\p)\subset (m^{(m)})$ 
and some function $u:[0,T]\times \mbb{R}^n\rightarrow \mbb{R}$ such that
$(u_{m^\p})$ converges to $u$ pointwise on the whole $[0,T]\times \mbb{R}^n$ space. 
Moreover, the function $u$ is actually continuous i.e. $u\in C([0,T]\times \mbb{R}^n)$. 
In fact, by the above construction of the sequence $(m^{(m)})$, 
$(u_{m^\p})$ converges uniformly to this function $u$ on any compact subset $\mbb{K}_R$.

In the remainder, we work on the sequence $(m^\p)$ (and possibly its further subsequences).
{\it Define} the \cadlag $\mbb{F}$-adapted process $(Y^{t,x}_s)_{s\in[0,T]}$ by
$Y^{t,x}_s:=u(s,X_s^{t,x}),\quad \forall (\omega,s)\in \Omega\times [0,T]$.
The uniform boundedness of $(u_{m^\p}, u)$, Lemma~\ref{lemma-x-continuity}(a) and  Chebyshev's inequality give 
\bea
||Y^{m^\p,t,x}-Y^{t,x}||^p_{\mbb{S}^p}\leq 
\mbb{E}\Bigl[\sup_{s\in[0,T]}\bigl| u_{m^\p}(s,X_s^{t,x})-u(s,X_s^{t,x})\bigr|^p\bold{1}_{\{\sup_{s\in[0,T]}|X_s^{t,x}|\leq R\}}\Bigr]
+C\Bigl(\frac{1+|x|^j}{R^j}\Bigr)~\nn
\eea
for every $R>0$ and $p, j\in\mbb{N}$ with some $m^\p$-independent constant $C$.  For a given $\ep>0$,
the 2nd term becomes smaller than $\ep/2$ with $R$ large enough.
Since $(u_{m^\p})$ converges uniformly to $u$ on any compact set, the first term also becomes smaller than $\ep/2$
for large $m^\prime$.  Hence, $||Y^{m^\p,t,x}-Y^{t,x}||^p_{\mbb{S}^p}\leq \ep$ for large $m^\p$.
Thus one concludes
$Y^{m^\p,t,x}\rightarrow Y^{t,x}$ in $\mbb{S}^p$ for every $p\in\mbb{N}$.
Since it implies $\sup_{s\in[0,T]}|Y^{m^\p,t,x}-Y_s^{t,x}|\rightarrow 0$ as $m^\p\rightarrow \infty$ in probability, 
extracting further subsequence (still denoted by $(m^\p)$), we have
$\lim_{m^\p\rightarrow \infty}\sup_{s\in[0,T]}|Y^{m^\p,t,x}_s-Y^{t,x}_s|=0$ $\mbb{P}$-a.s. 
In particular, it means $||Y^{m^\p,t,x}-Y^{t,x}||_{\mbb{S}^\infty}\rightarrow 0$.
It also implies that $(Y^{m^\prime,t,x})_{m^\prime}$ forms a Cauchy sequence in $\mbb{S}^\infty$.

With $m_1,m_2\in(m^\p)$, let us put $\del Y^{m_1,m_2}:=Y^{m_1,t,x}-Y^{m_2,t,x}$, $\del Z^{m_1,m_2}:=Z^{m_1,t,x}-Z^{m_2,t,x}$
and $\del \psi^{m_1,m_2}:=\psi^{m_1,t,x}-\psi^{m_2,t,x}$.  Ito 
formula applied to $|\del Y_t^{m_1,m_2}|^2$ yields for any $\tau\in\calt^T_0$,
\bea
&&\mbb{E}_{\calf_\tau}\left[|\del Y^{m_1,m_2}_\tau|^2+\int_\tau^T |\del Z^{m_1,m_2}_r|^2dr+\int_\tau^T\int_E |\del \psi^{m_1,m_2}_r(e)|^2 \mu(dr,de)\right] \nn \\
&&=\mbb{E}_{\calf_\tau}\left[\int_{\tau\vee t}^T 2\del Y^{m_1,m_2}_r \mbb{E}_{\calf_r}\Bigl[f_{m_1}(r,X_r^{t,x},
(Y_v^{m_1,t,x})_{v\in[r,T]},\Theta_r^{m_1,t,x})\right.\nn \\
&&\qquad \qquad \left.-f_{m_2}(r,X_r^{t,x},
(Y_v^{m_2,t,x})_{v\in[r,T]},\Theta_r^{m_2,t,x})\Bigr]dr\right]\nn
\eea
and hence
\bea
&&\mbb{E}_{\calf_\tau}\int_{\tau}^T |\del Z_r^{m_1,m_2}|^2 dr+\mbb{E}_{\calf_\tau}\int_\tau^T 
||\del \psi_r^{m_1,m_2}||^2_{\mbb{L}^2(\nu)}dr\nn \\
&&\leq 2||\del Y^{m_1,m_2}||_{\mbb{S}^\infty}\mbb{E}_{\calf_\tau}\int_\tau^T 
\sum_{i=1}^2|f_m(r,X_r^{t,x},(Y^{m_i,t,x}_v)_{v\in[r,T]},\Theta_r^{m_i,t,x})|dr~.\nn 
\eea
From Lemma~\ref{lemma-fm-regularized} and Assumption~\ref{assumption-LLC-M}, the conditional expectation of the 2nd line is bounded by $C\sum_{i=1}^2\bigl(1+||Y^{m_i,t,x}||_{\mbb{S}^\infty}
+||Z^{m_i,t,x}||^2_{\mbb{H}^2_{BMO}}+||\psi^{m_i,t,x}||^2_{\mbb{J}^2_{BMO}}\bigr)\leq C$, 
with $C=C(K_\cdot, A)$. Thus the right-hand side converges to zero as $m_1,m_2\rightarrow \infty$ uniformly in $\tau\in\calt_0^T$.
Therefore $\exists (Z^{t,x},\psi^{t,x})\in \mbb{H}^2_{BMO}\times \mbb{J}^2_{BMO}$ such that
$Z^{m^\p,t,x}\rightarrow Z^{t,x}$ in $\mbb{H}^2_{BMO}$ and $\psi^{m^\p,t,x}\rightarrow \psi^{t,x}$ in $\mbb{J}^2_{BMO}$.

Proving that $(Y^{t,x},Z^{t,x},\psi^{t,x})$ provides a solution of (\ref{eq-qg-bsde-M}) can be done via 
the common strategy for the BSDEs.
The above convergence results imply, a fortiori, that $Z^{m^\p,t,x}\rightarrow Z^{t,x}$ in $\mbb{H}^2$
and $\psi^{m^\p,t,x}\rightarrow \psi^{t,x}$ in $\mbb{J}^2$. Thus we also have the convergence in measure
for $Z^{m^\prime,t,x}\rightarrow Z^{t,x}$ and 
$\psi^{m^\prime,t,x}\rightarrow \psi^{t,x}$ with respect to $d\mbb{P}\otimes dt$ and
$d\mbb{P}\otimes \nu(de)\otimes dt$, respectively. As we have seen before, we also have 
$\sup_{s\in[0,T]}|Y^{m^\p,t,x}-Y_s^{t,x}|\rightarrow 0$ in probability.
By, for example, Corollary 6.13~\cite{Klenke} (treating general measure space with a $\sigma$-finite measure),  
there exists a subsequence (still denoted by $(m^\p)$) that yields almost everywhere convergence for the associated measure. 
Therefore, one has
$\sup_{s\in[0,T]}|Y^{m^\p,t,x}_s-Y^{t,x}_s|\rightarrow 0~{\rm a.s.}$, $Z^{m^\p,t,x}\rightarrow Z^{t,x}$ $d\mbb{P}\otimes ds$-a.e.
and  $\psi^{m,t,x}\rightarrow \psi^{t,x}$ $d\mbb{P}\otimes \nu(de)\otimes ds$-a.e.

Since $f_m\rightarrow f$ locally uniformly, the above a.e. convergences and the Lipschitz continuity of the driver
yields
\be
f_{m^\p}(s,X_s^{t,x},(Y^{m^\p,t,x}_v)_{v\in[s,T]}, \Theta^{m^\p,t,x}_s)
\rightarrow f(s,X_s^{t,x},(Y^{t,x}_v)_{v\in[s,T]},\Theta_s^{t,x})\nn
\ee
$d\mbb{P}\otimes ds$-a.e. 
In order to use the Lebesgue's dominated convergence theorem, we first show that
there exists an appropriate subsequence $(m^\p)$ such that
$G:=\sup_{m^\p}|Z^{m^\p,t,x}|^2$ and $H:=\sup_{m^\p}||\psi^{m^\p,t,x}||^2_{\mbb{L}^2(\nu)}$ are in $\mbb{L}^1(\Omega\times [0,T])$.
Let us follow the idea of Lemma 2.5 in \cite{Kobylanski}.
Since $(Z^{m^\prime,t,x})$ is a Cauchy sequence in $\mbb{H}^2$, one can extract a subsequence $(m^\prime_k)_{k\in\mbb{N}}$
such that for any $k\in \mbb{N}$,
$\displaystyle
||Z^{m^\p_{k+1},t,x}-Z^{m^\p_{k},t,x}||_{\mbb{H}^2}\leq 2^{-k}.\nn
$
On the other hand, for any $s\in[0,T]$, one easily sees that
\be
\displaystyle \sup_{k\in\mbb{N}}|Z^{m^\p_k,t,x}_s| \leq |Z^{m^\p_1,t,x}_s|+\sum_{k\in \mbb{N}}|Z^{m^\p_{k+1},t,x}_s-Z^{m^\p_k,t,x}_s|
.\nn
\ee
Taking the $\mbb{H}^2$-norm in the both side and using Minkowski's inequality,
\bea
\mbb{E}\Bigl[\int_0^T \sup_{k\in\mbb{N}}|Z^{m^\p_k,t,x}_s|^2 ds\Bigr]^\frac{1}{2} &\leq& ||Z^{m_1^\prime,t,x}||_{\mbb{H}^2}
+\sum_{k\in \mbb{N}}||Z^{m^\p_{k+1},t,x}-Z^{m^\p_k,t,x}||_{\mbb{H}^2}~\nn \\
&\leq & ||Z^{m_1^\prime,t,x}||_{\mbb{H}^2}+1<\infty~.\nn
\eea
Relabeling the subsequence by $(m^\p)$, one obtains the desired result for $G$.
Exactly the same method proves the integrability also for $H$.
Now, since $|f_{m^\p}|\leq C(1+G+H)$ a.s.with some $C=C(K_\cdot, A)$, we have
\bea
\int_0^T |f_{m^\p}(r,X_r^{t,x},(Y^{m^\p,t,x}_v)_{v\in[r,T]},\Theta^{m^\p,t,x}_r)-
f(r,X_r^{t,x},(Y^{t,x}_v)_{v\in[r,T]},\Theta_r^{t,x})|dr\rightarrow 0 ~{\rm a.s.}\nn
\eea
by the Lebesgue's dominated convergence theorem. 

Finally, the BDG inequality and the same arguments using the convergence in probability measure also give
$\sup_{s\in[0,T]}\Bigl|\int_s^T (Z_r^{m^\p,t,x}-Z_r^{t,x})dW_r\Bigr|\rightarrow 0~\rm{a.s.}$
and $\sup_{s\in[0,T]}\Bigl|\int_s^T\int_E (\psi_r^{m^\p,t,x}(e)-\psi_r^{t,x}(e))\wt{\mu}(dr,de)\Bigr|\rightarrow 0~\rm{a.s.}$
under appropriate subsequences, which guarantees the convergence for the stochastic integration.
This finishes the proof.
\end{proof}
\end{theorem}

\begin{remark}
By Theorem~\ref{th-existence-M} as well as the uniqueness of the solution, $Y_t^{t,x}$ is in fact deterministic in $(t,x)$.
\end{remark}

\begin{remark}
\label{remark-convergence}
In the above proof of Theorem~\ref{th-existence-M}, the convergence actually occurs in the entire sequence of $(m)$
not only the subsequence $(m^\p)$. If this is not the case, there must be a subsequence $(\hat{m})\subset (m)$
such that $||Y^{m_j,t,x}-Y^{t,x}||_{\mbb{S}^\infty}>c$ with some $c>0$ for every $m_j\in (\hat{m})$.
However, by repeating the same procedures done in the proof, we can extract a further subsequence 
$(\hat{m}^\p)\subset (\hat{m})$ such that,  $\exists (\wt{Y}^{t,x},\wt{Z}^{t,x},\wt{\psi}^{t,x})$,
$(Y^{m_j,t,x},Z^{m_j,t,x},\psi^{m_j,t,x})\rightarrow (\wt{Y}^{t,x},\wt{Z}^{t,x},\wt{\psi}^{t,x})$  in $\mbb{S}^\infty\times \mbb{H}^2_{BMO}\times \mbb{J}^2_{BMO}$ as 
$(\hat{m}^\p)\ni m_j\rightarrow \infty$.
One can show that it also provides the solution to (\ref{eq-qg-bsde-M}). By the uniqueness of solution, $\wt{Y}^{t,x}=Y^{t,x}$
in $\mbb{S}^\infty$, which contradicts the assumption.
\end{remark}

%%%%%%%%%%%%%%%%%%%%%%%%%%%%%%%%%
\section{Some regularity results}
%%%%%%%%%%%%%%%%%%%%%%%%%%%%%%%%
Due to the general path-dependence of $(Y_v)_{v\leq T}$ in the driver, it is difficult to establish
Malliavin's differentiability.
Interestingly, we can apply the method similar to Lemma 15 in Fromm \& Imkeller (2013)~\cite{Fromm-Imkeller}
or Lemma 2.5.14 in Fromm (2014)~\cite{Fromm} to derive some useful regularity results on the control variables.
The method only needs the fundamental Lebesgue's differentiation theorem.\footnote{See, for example, Section E.4, Theorem 6
\cite{Evans}.}

\begin{lemma}
Under Assumptions \ref{assumption-X}, \ref{assumption-structure-M} and \ref{assumption-LLC-M} with $\alpha=1$,
the control variables of the solution to the ABSDE (\ref{eq-qg-bsde-M}) satisfy the estimate for every $(t,x)$
\bea
|Z_{s}^{t,x}|\leq C\bigl(1+|X_{s}^{t,x}|^{1+\rho}\bigr), \quad ||\psi_s^{t,x}||_{\mbb{L}^2(\nu)}\leq C\bigl(1+|X_{s-}^{t,x}|^{1+\rho}\bigr) \nn
\eea
for $d\mbb{P}\otimes ds$-a.e. $(\omega,s)\in\Omega\times [0,T]$ with some constant $C=C(\rho,K_\xi,K, K_\cdot, A)$.
\begin{proof}
For notational simplicity, let us fix the initial data $(t,x)$ and omit the associated superscripts in the remainder of the proof.
We start from the regularized ABSDE (\ref{eq-absde-regularized}).
Choose any $\sp\in[0,T)$ and define $\del W_s:=W_s-W_{\sp}$ for $s\in[\sp,T]$.
An application of Ito formula to $(Y^m\del W^\top)$ yields 
\bea
Y_s^m\del W_s^\top&=&\int_{\sp}^s Z_r^m dr-\int_{\sp}^s\bold{1}_{r\geq t}\del W_r^\top\mbb{E}_{\calf_r}
f_m\bigl(r,X_r,(Y_v^m)_{v\in[r,T]},\Theta_r^m\bigr)dr\nn \\
&+&\int_{\sp}^s \del W_r^\top Z_r^m dW_r+\int_{\sp}^s\int_E \del W_r^\top \psi_r^m(e)\wt{\mu}(dr,de)
+\int_{\sp}^s Y_r^m dW_r^\top~.
\label{eq-Z-extract}
\eea
Since $(Y^m,Z^m,\psi^m)\in \mbb{S}^\infty\times \mbb{H}^2_{BMO}\times \mbb{J}^2_{BMO}$, one 
can show easily that the last three terms are true martingales. Notice that
\bea
&&\mbb{E}\Bigl[\int_0^T \bigl|\bold{1}_{r\geq t}W_r^\top \mbb{E}_{\calf_r}f_m\bigl(r,X_r,(Y^m_v)_{v\in[r,T]},\Theta_r^m\bigr)\bigr|dr\Bigr]\nn \\
&&\leq C\mbb{E}\Bigl[||W||^2_{[0,T]}\Bigr]^\frac{1}{2}\mbb{E}\Bigl[\Bigl(\int_0^T 
\bigl(1+|Z_r^m|^2+||\psi_r^m||^2_{\mbb{L}^2(\nu)}\bigr)dr\Bigr)^2\Bigr]^\frac{1}{2}\leq C \nn
\eea
with $C=C(K_\cdot, A)$. Thus Lebesgue's differentiation theorem implies that,  
\bea
&&\lim_{s\downarrow \sp}\frac{1}{s-\sp}\int_{\sp}^s\bold{1}_{r\geq t}W_r^\top
\mbb{E}_{\calf_r}f_m\bigl(r,X_r,(Y^m_v)_{v\in[r,T]},\Theta_r^m\bigr)dr\nn \\
&&\qquad =\bold{1}_{\sp\geq t}W_{\sp}^\top\mbb{E}_{\calf_{\sp}}f_m\bigl(\sp,X_{\sp},(Y^m_v)_{v\in[\sp,T]},\Theta_{\sp}^m\bigr)~~{\rm a.s.}\nn
\eea
for $dt$-a.e. $\sp\in[0,T)$. Similarly one obtains for $dt$-a.e. $\sp\in[0,T)$, 
\bea
&&\lim_{s\downarrow \sp}\frac{1}{s-\sp}\int_{\sp}^s Z_r^m dr=Z^m_\sp ~~{\rm a.s.} \nn \\
&&\lim_{s\downarrow \sp}\frac{1}{s-\sp}\int_{\sp}^s\bold{1}_{r\geq t}\mbb{E}_{\calf_r}f_m\bigl(r,X_r,(Y^m_v)_{v\in[r,T]},
\Theta_r^m\bigr)dr\nn \\
&&\qquad =\bold{1}_{\sp\geq t}\mbb{E}_{\calf_{\sp}}f_m\bigl(\sp,X_{\sp},(Y^m_v)_{v\in[\sp,T]},\Theta^m_{\sp}\bigr)dr~~{\rm a.s.}\nn
\eea
Since $Z^m\in \mbb{H}^2$, we can also take $\sp$ such that $\mbb{E}[|Z_{\sp}^m|]<\infty$ a.e. in $[0,T)$.

As in Lemma 2.5.14 of \cite{Fromm}, we introduce the stopping time $\tau:\Omega\rightarrow (\sp,T]$ such that
the following inequalities hold for all $s\in(\sp,T]$:
\bea
&&\bullet~ \Bigl|\frac{1}{s-\sp}\int_{\sp}^{\tau\wedge s}Z_r^mdr\Bigr|\leq |Z^m_{\sp}|+1~\quad {\rm a.s.} \nn\\
&&\bullet~ \Bigl|\frac{1}{s-\sp}\int_{\sp}^{\tau\wedge s}\bold{1}_{r\geq t}
\mbb{E}_{\calf_r}f_m\bigl(r,X_r,(Y^m_v)_{v\in[r,T]},\Theta_r^m\bigr)dr\Bigr| \nn \\
&&\qquad\qquad \leq\bold{1}_{\sp\geq t}\Bigl|\mbb{E}_{\calf_{\sp}}f_m
\bigl(\sp,X_{\sp},(Y^m_v)_{v\in[\sp,T]},\Theta_\sp^m\bigr)\Bigr|+1~\quad {\rm a.s.}  \nn\\
&&\bullet~ \Bigl|\frac{1}{s-\sp}\int_{\sp}^{\tau\wedge s}\bold{1}_{r\geq t}
W_r^\top \mbb{E}_{\calf_r}f_m\bigl(r,X_r,(Y^m_v)_{v\in[r,T]},\Theta_r^m\bigr)dr\Bigr| \nn \\
&&\qquad\qquad \leq\bold{1}_{\sp\geq t}\Bigl|W_{\sp}^\top \mbb{E}_{\calf_{\sp}}f_m
\bigl(\sp,X_{\sp},(Y^m_v)_{v\in[\sp,T]},\Theta_\sp^m\bigr)\Bigr|+1~\quad {\rm a.s.}~  \nn
\eea
Then one can show from (\ref{eq-Z-extract})
and the fact that $\tau(\omega) \wedge s=s$ for sufficiently small $s\in(\sp,T]$,
\bea
Z_{\sp}^m=\lim_{s\downarrow \sp}\mbb{E}_{\calf_\sp}\Bigl[\frac{1}{s-\sp}Y^m_{\tau\wedge s}(W_{\tau\wedge s}-W_\sp)^\top\Bigr]\nn
\eea
$d\mbb{P}\otimes dt$-a.e. $(\omega,\sp)\in\Omega\times [0,T)$ by the dominated convergence theorem. 
One sees
\bea
&&\Bigl|\mbb{E}_{\calf_{\sp}}\Bigl[\frac{1}{s-\sp}Y^m_{\tau\wedge s}(W_{\tau\wedge s}-W_{\sp})^\top\Bigr]\Bigr|\nn \\
&&\leq \Bigl|\mbb{E}_{\calf_\sp}\Bigl[\frac{1}{s-\sp}Y^m_s(W_{\tau\wedge s}-W_{\sp})^\top\Bigr]\Bigr|
+\Bigl|\mbb{E}_{\calf_\sp}\Bigl[\frac{1}{s-\sp}(Y^m_s-Y^m_{\tau\wedge s})(W_{\tau\wedge s}-W_{\sp})^\top\Bigr]\Bigr|\nn,
\eea
where the second term yields
\bea
&&\Bigl|\mbb{E}_{\calf_\sp}\Bigl[\frac{1}{s-\sp}(Y^m_s-Y^m_{\tau\wedge s})(W_{\tau\wedge s}-W_{\sp})^\top\Bigr]\Bigr|\nn \\
&&=\mbb{E}_{\calf_{\sp}}\Bigl[\frac{1}{s-\sp}\mbb{E}_{\calf_{\tau\wedge s}}\bigl[Y^m_s-Y^m_{\tau\wedge s}\bigr]
(W_{\tau\wedge s}-W_{\sp})^\top\Bigr]\Bigr|\nn \\
&&\leq \mbb{E}_{\calf_{\sp}}\Bigl[\frac{1}{s-\sp}\int_{\tau\wedge s}^s
\mbb{E}_{\calf_{\tau\wedge s}}\bigl|f_m(r,X_r,(Y^m_v)_{v\in[r,T]},\Theta_r^m)|dr (W_{\tau\wedge s}-W_{\sp})^\top\Bigr]\nn \\
&&\leq C_m\mbb{E}_{\calf_\sp}\Bigl[|W_{\tau\wedge s}-W_\sp|^2\Bigr]^\frac{1}{2}\leq C_m\sqrt{s-\sp}\rightarrow 0 \quad 
{\text{$s\downarrow \sp$}}. \nn
\eea
Here, we have used the fact that $|f_m|$ is essentially bounded for each $m$ (see Lemma~\ref{lemma-fm-regularized}).
The first term gives the estimate with some constant $C$ independent of $m$ such that
\bea
&&\Bigl|\mbb{E}_{\calf_\sp}\Bigl[\frac{1}{s-\sp}u_m(s,X_s)(W_{\tau\wedge s}-W_\sp)^\top\Bigr]\Bigr|=\Bigl|\mbb{E}_{\calf_\sp}\Bigl[\frac{1}{s-\sp}\bigl(u_m(s,X_s)-u_m(s,X_{\sp})\bigr)(W_{\tau\wedge s}-W_\sp)^\top\Bigr]\Bigr|\nn \\
&&\leq \frac{1}{\sqrt{s-\sp}}\mbb{E}_{\calf_{\sp}}\Bigl[|u_m(s,X_s)-u_m(s,X_\sp)|^2\Bigr]^\frac{1}{2}  \qquad\mbox{(by Cauchy-Schwartz)}
\nn\\
&&\leq \frac{C}{\sqrt{s-\sp}}\mbb{E}_{\calf_\sp}\Bigl[\bigl(1+[|X_s|\vee |X_{\sp}|]^{2\rho}\bigr)|X_s-X_{\sp}|^2\Bigr]^\frac{1}{2}
\qquad \mbox{(by (\ref{eq-um-continuity}) with $\alpha=1$)}
\nn \\
&&\leq \frac{C}{\sqrt{s-\sp}}\mbb{E}_{\calf_\sp}\Bigl[\bigl(1+|X_\sp|^{2\rho}+|X_s-X_\sp|^{2\rho}\bigr)|X_s-X_{\sp}|^2\Bigr]^\frac{1}{2}
~~\mbox{ (by $|x|\vee|y|\leq |x-y|+|y|$)}\\
&&\leq \frac{C}{\sqrt{s-\sp}}\Bigl\{(1+|X_\sp|^\rho)\mbb{E}_{\calf_\sp}\bigl[|X_s-X_{\sp}|^2]^\frac{1}{2} 
+\mbb{E}_{\calf_\sp}\bigl[|X_s-X_\sp|^{2(1+\rho)}\bigr]^\frac{1}{2}\Bigr\}
\nn \\
&&\leq C(1+|X_{\sp}|^{1+\rho}) \quad\mbox{a.s.}\nn
\eea
where, in the last inequality, we have used a conditional version of Lemma~\ref{lemma-x-continuity}(b)
with the initial value $X_{\sp}$. Thus we have $d\mbb{P}\otimes dt$-a.e.
\bea
|Z_{\sp}^m|\leq C(1+|X_{\sp}|^{1+\rho})  \nn
\eea
with $C=C(\rho,K_\xi,K,K_\cdot, A)$ uniformly in $m$. It is known from the proof of Theorem~\ref{th-existence-M} that $Z^m\rightarrow Z$ $d\mbb{P}\otimes dt$-a.e. under an appropriate subsequence, and hence the first claim follows.

The joint continuity of $u$ implies
$Y_{s-}=\lim_{r\uparrow s} u(r,X_r)=u(s,X_{s-})$ 
and hence 
\bea
\int_E|\psi_s(e)|^2\nu(de)&=&\int_E |u(s,X_{s-}+\gamma(s,X_{s-},e))-u(s,X_{s-})|^2\nu(de)\nn \\
&\leq & C \int_E\Bigl(1+|X_{s-}|^{2\rho}+|\gamma(s,X_{s-},e)|^{2\rho}\Bigr)|\gamma(s,X_{s-},e)|^2\nu(de)\nn \\
&\leq & C(1+|X_{s-}|^{2(1+\rho)})\int_E |e|^2\nu(de)\leq C(1+|X_{s-}|^{2(1+\rho)})~,\nn
\eea
which proves the second claim.
\end{proof}
\end{lemma}

%%%%%%%%%%%%%%%%%%%%%%%%%%%%%%%%%%%%%%%%%%%%%%%%%
\section{A non-Markovian setting}
\label{sec-non-Markovian}
%%%%%%%%%%%%%%%%%%%%%%%%%%%%%%%%%%%%%%%%%%%%%%%%%
\subsection{Existence}
%%%%%%%%%%%%%%%%%%%%%%%%%%%%%%%%%%%%%%%%%%%%%%%
In order to obtain the existence result in a non-Markovian setting, we need an additional so-called $A_\Gamma$-condition 
on the driver, which is rather restrictive but plays a crucial role in almost every existing 
work on quadratic growth BSDEs with jumps.
\begin{assumption}
\label{assumption-A-gamma}
For each $M>0$, for every $q\in \mbb{D}[0,T]$, $y\in\mbb{R}$, $z\in \mbb{R}^{1\times d}$, $\psi,\psi^\p\in \mbb{L}^2(E,\nu)$
with $\sup_{v\in[0,T]}|q_v|, |y|,||\psi||_{\mbb{L}^\infty(\nu)}, ||\psi^\p||_{\mbb{L}^\infty(\nu)}\leq M$
there exists a $\calp\otimes \cale$-measurable process $\Gamma^{q,y,z,\psi,\psi^\p, M}$ such that,
$d\mbb{P}\otimes dt$-a.e., 
\bea
f\bigl(t,(q_v)_{v\in[t,T]},y,z,\psi)-f(t,(q_v)_{v\in[t,T]},y,z,\psi^\prime)
\leq \int_E \Gamma_t^{q,y,z,\psi,\psi^\p,M}(e)\bigl(\psi(e)-\psi^\p(e)\bigr)\nu(de)\nn
\eea
with $C_M^1(1\wedge |e|)\leq \Gamma_t^{q,y,z,\psi,\psi^\p,M}(e)\leq C_M^2(1\wedge |e|)$
with two $M$ dependent constants satisfying $C_M^1>-1$ and $C_M^2\geq 0$.
\end{assumption}

We introduce a regularized ABSDE with some positive constant $m>0$:
\bea
Y^m_t&=&\xi+\int_t^T \mbb{E}_{\calf_r}f_m\bigl(r,(Y^m_v)_{v\in[r,T]},Y^m_r,Z^m_r,\psi^m_r\bigr)dr \nn \\
&&-\int_t^T Z_r^m dW_r-\int_t^T \int_E \psi^m_r(e)\wt{\mu}(dr,de), ~t\in[0,T]
\label{eq-bsde-regularized-nonM}
\eea
with the definition
$f_m\bigl(t,(q_v)_{v\in[t,T]},y,z,\psi\bigr):=f\bigl(t,(\varphi_m(q_v))_{v\in[t,T]},y,z,\psi\bigr)$
for every $(\omega,t,q,y,z,\psi)\in\Omega\times [0,T]\times \mbb{D}[0,T]\times \mbb{R}\times \mbb{R}^{1\times d}\times \mbb{L}^2(E,\nu)$.
$\varphi_m$ is the truncation function used previously. 
\begin{lemma}
\label{lemma-prelim-nonM}
If the driver $f$ satisfies Assumptions~\ref{assumption-structure-nonM}, \ref{assumption-LLC-nonM}
and \ref{assumption-A-gamma}, then the driver $f_m$ defined above also satisfies
the same conditions uniformly in $m$. Moreover, if there exists a bounded solution 
$(Y^m,Z^m,\psi^m)\in\mbb{S}^\infty\times \mbb{H}^2\times \mbb{J}^2$ to the ABSDE (\ref{eq-bsde-regularized-nonM}), 
then it is unique and  belongs
to $\mbb{S}^\infty\times \mbb{H}^2_{BMO}\times \mbb{J}^2_{BMO}$ with the norms
$||Y^m||_{\mbb{S}^\infty}, ||Z^m||_{\mbb{H}^2_{BMO}}, ||\psi^m||_{\mbb{J}^2_{BMO}}\leq C$ 
with some constant $C$ depending only on $A=(||\xi||_{\infty},||l||_{\mbb{S}^\infty},\del,\beta,\gamma, T)$.
\begin{proof}
The first claim is obvious. The second claim follows from Lemmas~\ref{lemma-z-psi-bound}, \ref{lemma-y-bound}
and Corollary~\ref{corollary-uniqueness-nonM}.
\end{proof}
\end{lemma}

\begin{theorem}
\label{theorem-existence-nonM}
Under Assumptions~\ref{assumption-structure-nonM}, \ref{assumption-LLC-nonM}
and \ref{assumption-A-gamma}, there exists a unique solution $(Y,Z,\psi)\in \mbb{S}^\infty\times \mbb{H}^2_{BMO}\times 
\mbb{J}^2_{BMO}$ to the ABSDE (\ref{eq-qg-bsde-nonM}).
\begin{proof}
Uniqueness follows from Corollary~\ref{corollary-uniqueness-nonM}.
Notice that it suffices to prove the existence of solution $(Y^{m},Z^m,\psi^m)\in \mbb{S}^\infty\times \mbb{H}^2_{BMO}\times \mbb{J}^2_{BMO}$
of (\ref{eq-bsde-regularized-nonM}) for each $m$. 
In fact, by choosing $m$ bigger than the bound given in Lemma~\ref{lemma-y-bound},
one sees $(Y^m,Z^m,\psi^m)$ actually provides the solution for (\ref{eq-qg-bsde-nonM}).
Let fix such an $m$ in the remainder.

Let us put $Y^{m,0}\equiv 0$ and define a sequence of BSDEs with $n\in\mbb{N}$ such that
\bea
&&Y^{m,n}_t=\xi+\int_t^T \mbb{E}_{\calf_r}f_m\bigl(r,(Y^{m,{n-1}}_v)_{v\in[r,T]},Y_r^{m,n},Z_r^{m,n},\psi_r^{m,n}\bigr)dr\nn \\
&&\quad -\int_t^T Z_r^{m,n}dW_r-\int_t^T\int_E \psi_r^{m,n}(e)\wt{\mu}(dr,de),~t\in[0,T]~.
\label{eq-bsde-ymn}
\eea
The driver for the BSDE (\ref{eq-bsde-ymn}) can be seen 
as $\wt{f}_m(r,y,z,\psi):=\mbb{E}_{\calf_r}f\bigl(r,(Y^{m,n-1}_v)_{v\in[r,T]},y,z,\psi\bigr)$.
By replacing $l_r$ by $l_r+\del m$, one sees the data $(\xi,\wt{f}_m)$ satisfy
Assumptions 3.1, 3.2 and 4.1 in \cite{FT-Qexp} for non-anticipated quadratic-exponential growth BSDEs.
Therefore, Theorem 4.1~\cite{FT-Qexp} implies that there exists a (unique) solution 
$(Y^{m,n},Z^{m,n},\psi^{m,n})\in \mbb{S}^\infty\times \mbb{H}^2_{BMO}\times \mbb{J}^2_{BMO}$
for each $n\geq 1$. Furthermore, as a special case of the universal bounds, one sees $||Y^{m,n}||_{\mbb{S}^\infty},||Z^{m,n}||_{\mbb{H}^2_{BMO}},
||\psi^{m,n}||_{\mbb{J}^2_{BMO}}\leq C$ with $C=C(||\xi||_{\infty},||l||_{\mbb{S}^\infty}+\del m, \beta,\gamma,T)$.

Let denote $\del Y^{m,n}:=Y^{m,n}-Y^{m,n-1}$.
Replacing $l_r$ by $l_r+\del m$, then putting $\del=0$, 
and considering the drivers $f^1(r,y,z,\psi):=f_m(r,(Y^{m,n}_v)_{v\in[r,T]},y,z,\psi)$, $\quad f^2(r,y,z,\psi):=f_m(r,(Y^{m,n-1}_v)_{v\in[r,T]},y,z,\psi)$,
one sees that $(f^i)_{i=1}^2$ satisfy Assumptions~\ref{assumption-structure-nonM} and \ref{assumption-LLC-nonM}.
Thus one can apply the stability results in Proposition~\ref{prop-stability} to the BSDE (\ref{eq-bsde-ymn}). 
In particular, by (\ref{eq-y-stability}), one has for any $p\geq 2q_*$ and $0<h\leq T$,
\bea
&&\mbb{E}\Bigl[\sup_{t\in[T-h,T]}|\del Y_t^{m,n+1}|^p\Bigr]\leq C\mbb{E}\Bigl[
\Bigl(\int_{T-h}^T \mbb{E}_{\calf_r}\bigl|f_m(r,(Y_v^{m,n})_{v\in[r,T]},\Theta_r^{m,n+1})\nn \\
&&\qquad -f_m(r,(Y_v^{m,n-1})_{v\in[r,T]},\Theta_r^{m,n+1})\bigr|dr\Bigr)^p\Bigr] 
\leq Ch^p\mbb{E}\Bigl[\sup_{t\in[T-h,T]}|\del Y_t^{m,n}|^p\Bigr]\nn
\eea 
with some constant $C=C(p,K_\cdot, ||\xi||_{\infty},||l||_{\mbb{S}^\infty}+\del m, \beta,\gamma,T)$.
By choosing $h$ small enough so that $Ch^p<1$, it becomes a strict contraction and thus $(Y^{m,n}_v, v\in[T-h,T])_{n\geq 1}$
forms a Cauchy sequence in $\mbb{S}^p[T-h,T]$.

By extracting an appropriate subsequence $(n^\p)\subset (n)$, one has $||\del Y^{m,n^\p}||_{\mbb{S}^\infty[T-h,T]}\rightarrow 0$
as $n^\p\rightarrow \infty$.
Applying Ito formula to $(\del Y^{m,n^\p})^2$  and repeating  
the same procedures used in last part of the proof in Theorem~\ref{th-existence-M},
one can show that $\exists (Y^m,Z^m,\psi^m) \in (\mbb{S}^\infty\times \mbb{H}^2_{BMO}\times 
\mbb{J}^2_{BMO})_{[T-h,T]}$, $(Y^{m,n^\p}, Z^{m,n^\p},\psi^{m,n^\p})\rightarrow (Y^m,Z^m,S^m)$
in the corresponding norm, and that $(Y^m_v,Z^m_v,\psi^m_v)_{v\in[T-h,T]}$ solves the ABSDE (\ref{eq-bsde-regularized-nonM}) for 
the period $[T-h,T]$.\footnote{Thanks to the uniqueness of the solution of (\ref{eq-bsde-regularized-nonM}), the same arguments
used in Remark~\ref{remark-convergence} guarantee that the above convergence actually occurs in the entire sequence $(n)$. }

Now, let us replace $(Y^{m,n},Z^{m,n},\psi^{m,n})_{n\in\mbb{N}}$ by $(Y^{m},Z^m,\psi^m)$ for $(\omega,s)\in\Omega\times [T-h,T]$
in (\ref{eq-bsde-ymn}). Then for $t\leq T-h$, we have
\bea
&&Y_t^{m,n}=Y^m_{T-h}+\int_t^{T-h} \mbb{E}_{\calf_r}f_m\bigl(r,(Y^{m,{n-1}}_v)_{v\in[r,T]},Y_r^{m,n},Z_r^{m,n},\psi_r^{m,n}\bigr)dr\nn \\
&&\quad -\int_t^{T-h} Z_r^{m,n}dW_r-\int_t^{T-h}\int_E \psi_r^{m,n}(e)\wt{\mu}(dr,de).\nn
\eea
An application of Proposition~\ref{prop-stability} with the data $(Y^m_{T-h},f^1), (Y^m_{T-h},f^2)$ yields, 
\bea
\mbb{E}\Bigl[\sup_{t\in[T-2h,T-h]}|\del Y^{m,n+1}|^p\Bigr]&\leq& C\mbb{E}\Bigl[\Bigl(\int_{T-2h}^{T-h}\mbb{E}_{\calf_r}|\del f(r)|dr\Bigr)^p\Bigr]\nn \\
&\leq & C h^p\mbb{E}\Bigl[\sup_{t\in[T-2h,T]}|\del Y_t^{m,n}|^p\Bigr]
=Ch^p\mbb{E}\Bigl[\sup_{t\in[T-2h,T-h]}|\del Y_t^{m,n}|^p\Bigr]\nn
\eea
where the fact  $Y^{m,n}_s=Y^m_s$, $s\in[T-h,T]$ is used in the 2nd line. Thus one can extend the solution to the period $[T-2h,T-h]$
by the same procedures used in the previous step. Since coefficient $C$ can be taken independently of the specific period,
the whole period $[0,T]$ can be covered by a finite number of partitions.
Notice here that, as one can see from the proof of Proposition~\ref{prop-stability}, the coefficient $C$ depends on the essential supremum of
the terminal value $||\xi||_{\infty}$ only through the local Lipschitz constant $K_M$ and the universal bounds controlling $M$ as well as
the coefficients of the reverse H\"older inequality.
Hence the appearance of the new terminal value $Y^{m}_{T-h}$ does not change the size of the coefficient $C$.
This finishes the proof for the existence of a bounded solution to (\ref{eq-bsde-regularized-nonM}) for each $m$.
\end{proof}
\end{theorem}

%%%%%%%%%%%%%%%%%%%%%%%%%%%%%%%%%%%%%
\subsection{Comparison principle}
%%%%%%%%%%%%%%%%%%%%%%%%%%%%%%%%%%%%%
For completeness, we give a sufficient condition for the comparison principle to hold for our ABSDE
in the rest of this section.
In non-anticipated settings, i.e. when there is no future path-dependence of $(Y_v)_{v\in[0,T]}$ in the driver $f$,
it is known  that the comparison principle holds for quadratic-exponential growth BSDEs in the presence of 
$A_\Gamma$-condition (See, Lemma~\ref{lemma-comparison-nonA}.). For the current anticipated setting, we need an additional assumption 
same as the one used in Theorem 5.1 of \cite{Peng-Yang}.
Consider the two ABSDEs with $i\in\{1,2\}$,
\bea
Y_t^i=\xi_i+\int_t^T \mbb{E}_{\calf_r}f_i\bigl(r,(Y_v^i)_{v\in[r,T]},Y_r^i,Z_r^i,\psi_r^i\bigr)dr
-\int_t^T Z_r^i dW_r-\int_t^T \int_E \psi_r^i(e)\wt{\mu}(dr,de)\nn
\eea
for $t\in[0,T]$.
\begin{theorem}
Suppose the data $(\xi_i,f_i)_{1\leq i\leq 2}$ satisfy Assumptions~\ref{assumption-structure-nonM},
\ref{assumption-LLC-nonM} and \ref{assumption-A-gamma}.
Moreover, $f_2$ is increasing in $(q_v)_{v\in[0,T]}$, i.e. $f_2(r,(q_v)_{v\in[r,T]},y,z,\psi)
\leq f_2(r,(q^\prime_v)_{v\in[r,T]},y,z,\psi)$ 
for every $(r,y,z,\psi)\in[0,T]\times \mbb{R}\times \mbb{R}^{1\times d}\times \mbb{L}^2(E,\nu)$ 
and $q,q^\prime\in\mbb{D}[0,T]$, if $q_v\leq q^\prime_v$ $\forall v\in[r,T]$.
If $\xi_1\leq \xi_2$ a.s. and $f_1(r,(q_v)_{v\in[r,T]},y,z,\psi)\leq f_2(r,(q_v)_{v\in[r,T]},y,z,\psi)$
$d\mbb{P}\otimes dr$-a.e. for every $(q,y,z,\psi)\in \mbb{D}[0,T]\times \mbb{R}\times \mbb{R}^{1\times d}\times \mbb{L}^2(E,\nu)$,
then $Y^1_t\leq Y^2_t~\forall t\in[0,T]$ a.s.
\begin{proof}
Firstly, let us regularize the driver $f_2$ by $f_2^\prime$ defined as, for every $(r,q,y,z,\psi)$, 
\bea
f_2^\p(r,(q_v)_{r\in[t,T]},y,z,\psi):=f_2\bigl(r,(\varphi_m(q_v))_{v\in[r,T]},y,z,\psi\bigr)
\eea
with some truncation level $m$ satisfying $m>(||Y^1||_{\mbb{S}^\infty}\vee ||Y^2||_{\mbb{S}^\infty})$. Consider a sequence of non-anticipated BSDEs with $n\in\mbb{N}$ by
\bea
Y_t^{2,n}&=&\xi_2+\int_t^T \mbb{E}_{\calf_r}f_2^\p\bigl(r,(Y_v^{2,n-1})_{v\in[r,T]},Y_r^{2,n},Z_r^{2,n},\psi_r^{2,n}\bigr)dr\nn \\
&&-\int_t^T Z_r^{2,n}dW_r-\int_t^T \int_E \psi_r^{2,n}(e)\wt{\mu}(dr,de),\quad t\in[0,T]
\label{eq-comp-sequence}
\eea
under the condition $Y^{2,0}=Y^1$.
By the proof of Theorem~\ref{theorem-existence-nonM}, there exists $h>0$ such that
$(Y^{2,n},Z^{2,n},\psi^{2,n})\rightarrow (Y^2,Z^2,\psi^2)$ in $\mbb{S}^\infty\times \mbb{H}^2_{BMO}\times \mbb{J}^2_{BMO}$
as $n\rightarrow \infty$ for the period $[T-h,T]$. Note that the constraint $\varphi_m(\cdot)$
becomes passive at least for large enough $n$.

Firstly, let us focus on the period $[T-h,T]$.
Set $\wt{f}_1(r,y,z,\psi)=\mbb{E}_{\calf_r}f_1(r,(Y_v^1)_{v\in[r,T]},y,z,\psi)$
and $\wt{f}_2(r,y,z,\psi)=\mbb{E}_{\calf_r}f_2^\p(r,(Y_v^1)_{v\in[r,T]},y,z,\psi)=\mbb{E}_{\calf_r}f_2(r,(Y_v^1)_{v\in[r,T]},y,z,\psi)$. 
Applying Lemma~\ref{lemma-comparison-nonA}, one obtains $Y^1_t=Y^{2,0}_t\leq Y^{2,1}_t~\forall t\in[T-h,T]$ a.s.
Then using the new definition
\bea
&&\wt{f}_1(r,y,z,\psi)=\mbb{E}_{\calf_r}f_2^\p(r,(Y_v^{2,0})_{v\in[r,T]},y,z,\psi), \nn \\
&&\wt{f}_2(r,y,z,\psi)=\mbb{E}_{\calf_r}f_2^\p(r,(Y_v^{2,1})_{v\in[r,T]},y,z,\psi), \nn
\eea
and the hypothesis that the driver is increasing in $q\in\mbb{D}[0,T]$,
Lemma~\ref{lemma-comparison-nonA} yields $Y^{2,1}_t\leq Y^{2,2}~\forall t\in[T-h,T]$ a.s. 
By repeating the same arguments, one sees $Y^1_t\leq Y^{2,n-1}_t\leq Y^{2,n}_t~\forall t\in[T-h,T]$ a.s.
for every $n\in\mbb{N}$.
Since $Y^{2,n}$ converges to $Y^2$ in $\mbb{S}^\infty[T-h,T]$, 
one concludes $Y^1_t\leq Y^2_t~\forall t\in[T-h,T]$ a.s.

Let us now replace $Y^{2,n}_t$ by $Y^2_t$ for all $t\in[T-h,T]$ in (\ref{eq-comp-sequence}),
and consider a sequence of non-anticipated BSDEs $n\in\mbb{N}$
\bea
Y^{2,n}_t&=&Y^2_{T-h}+\int_t^{T-h}\mbb{E}_{\calf_r}f_2^\p\bigl(r,(Y_v^{2,n-1})_{v\in[r,T]},Y_r^{2,n},Z_r^{2,n},\psi_r^{2,n}\bigr)dr\nn \\
&&-\int_t^{T-h}Z_r^{2,n}dW_r-\int_t^{T-h}\int_E \psi_r^{2,n}(e)\wt{\mu}(dr,de) \nn
\eea
with the initial condition 
$Y_t^{2,0}=\begin{cases} Y_t^1,~t\in[0,T-h)\\
Y_t^2,~t\in[T-h,T]
\end{cases}$ for the next short period $t\in[T-2h,T-h]$.
By the result of the previous step, one has $Y^1_t\leq Y^{2,0}_t~\forall t\in[T-2h,T]$ a.s.
Now, let us set $\wt{f}_1(r,y,z,\psi)=\mbb{E}_{\calf_r}f_1(r,(Y^1_v)_{v\in[r,T]},y,z,\psi)$, $\wt{f}_2(r,y,z,\psi)=\mbb{E}_{\calf_r}f_2^\prime(r,(Y^{2,0}_v)_{v\in[r,T]},y,z,\psi)$,
where the latter is equal to $\mbb{E}_{\calf_r}f_2(r,(Y^{2,0}_v)_{v\in[r,T]},y,z,\psi)$.
By applying Lemma~\ref{lemma-comparison-nonA} to the data $(Y^1_{T-h},\wt{f}_1)$, $(Y^2_{T-h},\wt{f}_2)$,
one obtains $Y^1_t\leq Y^{2,1}_t~\forall t\in[T-2h,T-h]$ a.s.
Since $Y^{2,1}_t=Y^2_t$ for $t\in[T-h,T]$, one concludes $Y^1_t\leq Y^{2,0}_t\leq Y^{2,1}_t~\forall t\in[T-2h,T]$ a.s.
Similarly, applying Lemma~\ref{lemma-comparison-nonA} with $\wt{f}_1(r,y,z,\psi)=\mbb{E}_{\calf_r}f_2^\p(r,(Y^{2,n-2}_v)_{v\in[r,T]},
y,z,\psi)$, $\wt{f}_2(r,y,z,\psi)=\mbb{E}_{\calf_r}f_2^\p(r,(Y^{2,n-1}_v)_{v\in[r,T]},y,z,\psi)$ yields
$Y^{2,n-1}_t\leq Y^{2,n}_t~\forall t\in[T-2h,T]$ a.s. for every $n\geq 2$.
As in the previous step, the proof of Theorem~\ref{theorem-existence-nonM} implies 
$Y^{2,n}\rightarrow Y^2$ in $\mbb{S}^\infty[T-2h,T-h]$. Since $Y^{2,n}_t=Y^{2}_t$ for $t\in[T-h,T]$ by construction,
one actually has $Y^{2,n}\rightarrow Y^2$ in $\mbb{S}^\infty[T-2h,T]$.
It follows that $Y^1_t\leq Y^2_t~\forall t\in[T-2h,T]$ a.s.  
Repeating the same procedures finite number of times, one obtains the desired result.
\end{proof}
\end{theorem}

%%%%%%%%%%%%%%%%%%%%%%%%%%%%%%%%%%%%%%%%%%%%%%%%%%%%%%%%%%%%%%%%%%%%%%%%%%%%%%%%%%%%%%%%%%%%%%%%%%%%%%%%%%%%%%%%%%%%%%%%%%%%%%%
%
%%%%%%%%%%%%%%%          Appendix
%%%%%%%%%%%%%%%
%
%%%%%%%%%%%%%%%%%%%%%%%%%%%%%%%%%%%%%%%%%%%%%%%%%%%%%%%%%%%%%%%%%%%%%%%%%%%%%%%%%%%%%%%%%%%%%%%%%%%%%%%%%%%%%%%%%%%%%%%%%%%%%%%%
\begin{appendix}
\section{Some preliminary results}
Let us remind some important properties of BMO-martingales.
For our purpose, it is enough to focus on continuous ones.
When $Z\in \mbb{H}^2_{BMO}$, $M_\cdot:=\int_0^\cdot Z_rdW_r$ is a continuous BMO-martingale with
$||M||_{BMO}=||Z||_{\mbb{H}^2_{BMO}}$.
\begin{lemma} [reverse H\"older inequality]
\label{lemma-R-Holder}
Let $M$ be a continuous BMO-martingale.
Then,  Dol\'eans-Dade exponential $\bigl(\cale_t(M),t\in[0,T]\bigr)$ is a uniformly integrable martingale,  and 
for every stopping time $\tau\in\calt^T_0$, there exists some $r>1$
such that $
\mbb{E}\left[\cale_T(M)^r|\calf_\tau\right]\leq C \cale_\tau (M)^r$
with some positive constant $C=C(r,||M||_{BMO})$.
\begin{proof}
See Kazamaki (1979)~\cite{Kazamaki}, and also Remark 3.1 of Kazamaki (1994)~\cite{Kazamaki-note}.
\end{proof}
\end{lemma}

\begin{lemma}
\label{lemma-BMO-PQ}
Let $M$ be a square integrable continuous martingale and $\hat{M}:=\langle M \rangle -M$.
Then, $M\in BMO(\mbb{P})$ if and only if $\hat{M}\in BMO(\mbb{Q})$
with $d\mbb{Q}/d\mbb{P}=\cale_T(M)$. Furthermore, $||\hat{M}||_{BMO(\mbb{Q})}$ is 
determined by some function of $||M||_{BMO(\mbb{P})}$ and vice versa.
\begin{proof}
See Theorem 3.3 and Theorem 2.4 in \cite{Kazamaki-note}.
\end{proof}
\end{lemma}

\begin{remark}
\label{remark-BMO}
For continuous martingales, Theorem 3.1~\cite{Kazamaki-note} also tells that
there exists some decreasing function $\Phi(r)$ with $\Phi(1+)=\infty$
and $\Phi(\infty)=0$ such that if $||M||_{BMO(\mbb{P})}$ satisfies
$ ||M||_{BMO(\mbb{P})}<\Phi(r)$
then $\cale(M)$ satisfies the reverse H\"older inequality with power $r$.
This implies together with Lemma~\ref{lemma-BMO-PQ}, one can take 
a common positive constant $\bar{r}$ satisfying $1<\bar{r}\leq r^*$ such that both of 
the $\cale(M)$ and $\cale(\hat{M})$ satisfy the reverse H\"older inequality
with power $\bar{r}$ under the respective probability measure $\mbb{P}$ and $\mbb{Q}$.
Furthermore,  the upper bound $r^*$ is determined only by $||M||_{BMO(\mbb{P})}$ (or equivalently by $||M||_{BMO(\mbb{Q})}$).
\end{remark}

Let us also remind the following result.
\begin{lemma}{(Chapter 1, Section 9, Lemma 6~\cite{Shiryayev})}
\label{lemma-Shiryayev}
For any $\Psi\in \mbb{J}^p$ with $p\geq 2$, there exists some constant $C=C(p)$ such that
\bea
\mbb{E}\Bigl[\Bigl(\int_0^T\int_E |\Psi_r(e)|^2\nu(de)dr\Bigr)^\frac{p}{2}\Bigr]
\leq C\mbb{E}\Bigl[\Bigl(\int_0^T\int_E |\Psi_r(e)|^2\mu(dr,de)\Bigr)^\frac{p}{2}\Bigr]~.\nn
\eea
\end{lemma}

\begin{lemma} (Lemma 5-1 of Bichteler, Gravereaux and Jacod (1987)~\cite{BGJ})
\label{ap-lemma-BGJ}
Let $\eta:\mbb{R}\rightarrow \mbb{R}$ be defined by $\eta(e)=1\wedge |e|$.
Then, for $\forall p\geq 2$, there exists a constant $\del_p$ depending on $p,T,n,k$ 
such that
\bea
\mbb{E}\left[\sup_{t\in[0,T]}\Bigl|\int_0^t\int_E U(s,e)\wt{\mu}(ds,de)\Bigr|^p\right]
\leq \del_p \int_0^T \mbb{E}|L_s|^p ds \nn
\eea 
if $U$ is an $\mbb{R}^{n\times k}$-valued $\calp\otimes \cale$-measurable function on $\Omega\times[0,T]\times E$
and $L$ is a predictable process satisfying $|U_{\cdot}^i(\omega,s,e)|\leq L_s(\omega)\eta(e)$ for
each column $1\leq i\leq k$.
\end{lemma}

%%%%%%%%%%%%%%%%%%%%%%%%%%%%%%%%%%%%%%%%%%%%%%%%%%%%%%%%%%
\section{Technical details omitted in the main text}
%%%%%%%%%%%%%%%%%%%%%%%%%%%%%%%%%%%%%%%%%%%%%%%%%%%%%%%%%%
In the main text, we have omitted some technical details in order not to interrupt the 
main story. In this section, let us give the omitted details for completeness.
%%%%%%%%%%%%%%%%%%%%%%%%%%%%%%%%%%%%%%%%%%%%%%%%%%%%%%%%%%%%%%%%%%%%%%%%
\subsection{Details of the proof of Lemma~\ref{lemma-z-psi-bound}}
\label{ap-L3-1}
%%%%%%%%%%%%%%%%%%%%%%%%%%%%%%%%%%%%%%%%%%%%%%%%%%%%%%%%%%%%%%%%%%%%%%%%%
By assumption, we have $Y\in\mbb{S}^\infty$. Since $||\psi||_{\mbb{J}^\infty}\leq 2||Y||_{\mbb{S}^\infty}$,
$\psi$ is bounded.  Ito formula applied to $e^{2\gamma Y_t}$ yields, for any $\mbb{F}$-stopping time $\tau\in\calt_0^T$,
\bea
&&\mbb{E}_{\calf_\tau} \left[
\int_\tau^T e^{2\gamma Y_s}2\gamma^2 |Z_s|^2 ds+\int_\tau^T \int_E  e^{2\gamma Y_s} (e^{\gamma \psi_s(e)}-1)^2\nu(de)ds\right]\nn \\
&&=\mbb{E}_{\calf_\tau}\left[e^{2\gamma Y_T}-e^{2\gamma Y_{\tau}}+2\gamma \int_{\tau}^T e^{2 \gamma Y_s}
\Bigl(\mbb{E}_{\calf_s}f\bigl(s,(Y_v)_{v\in[s,T]},Y_s,Z_s,\psi_s\bigr)-\int_E j_\gamma(\psi_s(e))\nu(de)\Bigr)ds\right]\nn \\
&&\leq \mbb{E}_{\calf_\tau}\left[
e^{2\gamma Y_T}-e^{2\gamma Y_\tau}+2\gamma \int_\tau^Te^{2\gamma Y_s}\Bigl(l_s+ \del ||Y||_{[s,T]}+\beta |Y_s|+\frac{\gamma}{2}|Z_s|^2\Bigr)ds
\right]\nn
\eea
where structure condition in Assumption~\ref{assumption-structure-nonM} was used in the third line. Then it yields
\bea
&&\mbb{E}_{\calf_\tau}\left[\int_\tau^T e^{2\gamma Y_s}\gamma^2 |Z_s|^2 ds+\int_\tau^T \int_E  e^{2\gamma Y_s} (e^{\gamma \psi_s(e)}-1)^2\nu(de)ds\right]\nn \\
&&\leq e^{2\gamma ||Y||_{\mbb{S}^\infty}}+2\gamma e^{2\gamma ||Y||_{\mbb{S}^\infty}}T\Bigl(
||l||_{\mbb{S}^\infty}+(\beta+\del)||Y||_{\mbb{S}^\infty}\Bigr). \nn
\eea
Since $e^{-2\gamma ||Y||_{\mbb{S}^\infty}}\leq e^{\pm 2\gamma Y}\leq e^{2\gamma ||Y||_{\mbb{S}^\infty}}$,
\bea
&&\mbb{E}_{\calf_\tau}\left[\int_\tau^T \gamma^2 |Z_s|^2 ds+\int_\tau^T \int_E(e^{\gamma \psi_s(e)}-1)^2\nu(de)ds\right]\nn \\
&&\leq e^{4\gamma ||Y||_{\mbb{S}^\infty}}+2\gamma e^{4\gamma ||Y||_{\mbb{S}^\infty}}T\Bigl(
||l||_{\mbb{S}^\infty}+(\beta+\del)||Y||_{\mbb{S}^\infty}\Bigr). \nn
\eea
In particular, this leads to the desired bound on $||Z||^2_{\mbb{H}^2_{BMO}}$.

Repeating the same calculation on $e^{-2\gamma Y_t}$, one obtains the next estimate:
\bea
&&\mbb{E}_{\calf_\tau}\left[\int_\tau^T \gamma^2 |Z_s|^2 ds+\int_\tau^T \int_E(e^{-\gamma \psi_s(e)}-1)^2\nu(de)ds\right]\nn \\
&&\leq e^{4\gamma ||Y||_{\mbb{S}^\infty}}+2\gamma e^{4\gamma ||Y||_{\mbb{S}^\infty}}T\Bigl(
||l||_{\mbb{S}^\infty}+(\beta+\del)||Y||_{\mbb{S}^\infty}\Bigr). \nn
\eea
Noticing the fact that $(e^x-1)^2+(e^{-x}-1)^2\geq x^2, \forall x\in \mbb{R}$, one obtains
\bea
||\psi||^2_{\mbb{J}^2_B}\leq \frac{e^{4\gamma ||Y||_{\mbb{S}^\infty}}}{\gamma^2}
\Bigl(2+4\gamma T \bigl(||l||_{\mbb{S}^\infty}+(\beta+\gamma)||Y||_{\mbb{S}^\infty}\bigr)\Bigr).\nn
\eea
Finally, the relation
\be
||\psi||^2_{\mbb{J}^2_{BMO}}\leq ||\psi||^2_{\mbb{J}^2_B}+||\psi||^2_{\mbb{J}^\infty}
\leq ||\psi||^2_{\mbb{J}^2_B}+4||Y||^2_{\mbb{S}^\infty}\nn
\ee
proves the desired estimate on $||\psi||^2_{\mbb{J}^2_{BMO}}$.

%%%%%%%%%%%%%%%%%%%%%%%%%%%%%%%%%%%%%%%%%%%%%%%%%%%%%%
\subsection{Derivation of (\ref{eq-dP})}
\label{ap-L3-2}
%%%%%%%%%%%%%%%%%%%%%%%%%%%%%%%%%%%%%%%%%%%%%%%%%%%%%%%
Using (\ref{eq-d-group}), one gets
\bea
&&dP_t=P_{t-}\gamma d\Bigl(e^{\beta t}|Y_t|+\int_0^t e^{\beta r}\bigl(l_r+\del \mbb{E}_{\calf_r}\sup_{v\in[r,T]}|Y_v|\bigr)dr\Bigr)
+P_t\frac{\gamma^2}{2}|e^{\beta t}\sign(Y_t)Z_t|^2 dt\nn \\
&&\qquad +P_{t-}\int_E \Bigl(e^{\gamma e^{\beta t}(|Y_{t-}+\psi_t(e)|-|Y_{t-}|)}-1-\gamma e^{\beta t}\sign(Y_{t-})\psi_t(e)\Bigr)\mu(dt,de)\nn \\
&&=P_{t-}\int_E \Bigl(e^{\gamma e^{\beta t}(|Y_{t-}+\psi_t(e)|-|Y_{t-}|)}-1-\gamma e^{\beta t}\sign(Y_{t-})\psi_t(e)\Bigr)\mu(dt,de)\nn \\
&&\qquad +P_t\frac{\gamma^2}{2}|e^{\beta t}\sign(Y_t)Z_t|^2 dt+P_{t-}\gamma\left\{ 
e^{\beta t}\sign(Y_t)Z_t dW_t+\int_E e^{\beta t}\sign(Y_{t-})\psi_t(e)\wt{\mu}(dt,de)
\right.\nn \\
&&\qquad \left.-\int_E j_\gamma\bigl(e^{\beta t}\sign(Y_t)\psi_t(e)\bigr)\nu(de)dt-\frac{\gamma}{2}|e^{\beta t}\sign(Y_t)Z_t|^2dt
+dC_t\right\}.\nn
\eea
Separating the terms contained in $dC^\prime$ (\ref{eq-dCprime}) and canceling $|Z|^2$-term, one obtains
\bea
&&dP_t=P_{t-}\left\{
\gamma dC_t+\int_E \Bigl(e^{\gamma e^{\beta t}(|Y_{t-}+\psi_t(e)|-|Y_{t-}|)}-e^{\gamma e^{\beta t} \sign(Y_{t-})\psi_t(e)}\Bigr)\mu(dt,de)
\right\}\nn \\
&&+P_{t-}\int_E\Bigl(e^{\gamma e^{\beta t}\sign(Y_{t-})\psi_t(e)}-1-\gamma e^{\beta t}\sign(Y_{t-})\psi_t(e)\Bigr)\mu(dt,de)\nn \\
&&+P_{t-}\gamma\left\{
e^{\beta t}\sign(Y_{t})Z_tdW_t+\int_E e^{\beta t}\sign(Y_{t-})\psi_t(e)\wt{\mu}(dt,de)
-\int_E j_\gamma\bigl(e^{\beta t}\sign(Y_t)\psi_t(e)\bigr)\nu(de)dt\right\}~.
\nn
\eea
Notice that the terms inside a parenthesis in the second line are equal to $\gamma j_\gamma \bigl(e^{\beta t}\sign(Y_{t-})\psi_t(e)\bigr)$,
which then yields
\bea
&& dP_t=P_{t-}dC_t^\prime+P_{t-}\int_E \gamma j_{\gamma}\bigl(e^{\beta t}\sign(Y_{t-})\psi_t(e)\bigr)\wt{\mu}(dt,de)\nn \\
&&\qquad +P_{t-}\left\{
\gamma e^{\beta t}\sign(Y_t)Z_t dW_t+\int_E \gamma e^{\beta t}\sign(Y_{t-})\psi_t(e)\wt{\mu}(dt,de)\right\}\nn.
\eea
Using the definition of $j_\gamma(\cdot)$, one obtains the desired expression (\ref{eq-dP}).

%%%%%%%%%%%%%%%%%%%%%%%%%%%%%%%%%%%%%%%%%%%%%%%%%%%%%%
\subsection{The proof for Lemma~\ref{lemma-x-continuity}}
\label{ap-x-continuity}
%%%%%%%%%%%%%%%%%%%%%%%%%%%%%%%%%%%%%%%%%%%%%%%%%%%%%%%
The existence of unique solution $X^{t,x}\in \mbb{S}^p,~\forall p\geq 2$ and $\forall (t,x)\in[0,T]\times \mbb{R}^n$
is well known for the Lipschitz SDEs with jumps.  Hence, we only provide a proof for the 
relevant continuities below.\\
(a) For any $s\in[t,T]$ and $p\geq 2$, the BDG inequality yields
\bea
&&\mbb{E}[|X_s^{t,x}|^p]\leq C\mbb{E}\Bigl\{|x|^p+\Bigl(\int_t^s |b(r,X_r^{t,x})|dr\Bigr)^p
+\Bigl(\int_t^s |\sigma(r,X_r^{t,x})|^2dr\Bigr)^\frac{p}{2}\nn \\
&&\qquad \qquad +\sup_{u\in[t,s]}\Bigl|\int_t^u\int_E |\gamma(r,X_{r-}^{t,x},e)|\wt{\mu}(dr,de)\Bigr|^p\Bigr\}
\nn
\eea
Since for each $1\leq i\leq k$, we have $|\gamma^i(r,X_{r-}^{t,x},e)|\leq K(1+|X_{r-}^{t,x}|)\eta(e)$ 
by Assumption~\ref{assumption-X} (ii) and (iii).
By Lemma~\ref{ap-lemma-BGJ} and the Lipschitz continuity yields,
\bea
\mbb{E}[|X_s^{t,x}|]\leq C(1+|x|^p)+C\int_t^s \mbb{E}[|X_r^{t,x}|^p]dr \nn
\eea
and hence the Gronwall inequality gives $\sup_{s\in[t,T]}\mbb{E}[|X_s^{t,x}|^p]\leq C(1+|x|^p)$.
Noticing the fact that $X_s^{t,x}\equiv x$ for $s\leq t$ and applying BDG inequality once again,
one obtains 
\bea
\mbb{E}\Bigl[\sup_{s\in[0,T]}|X_s^{t,x}|^p\Bigr]\leq C(1+|x|^p)~.\nn
\eea
%%%%%%%%%%%%%%%%%%%%%%%%%%%%%%%%%%
(b) Let us assume $t\leq s\leq u\leq s+h$. The case with $s<t$ can be done similarly by 
using $X_s^{t,x}\equiv x$ for $s\leq t$.
Since
\bea
X_u^{t,x}-X_s^{t,x}=\int_s^u b(r,X_r^{t,x})dr+\int_s^u\sigma(r,X_r^{t,x})dW_r+\int_s^u\int_E \gamma(r,X_{r-}^{t,x})\wt{\mu}(dr,de)~.\nn
\eea
Using the BDG inequality, Lemma~\ref{ap-lemma-BGJ} and the result (a), one obtains
\bea
\mbb{E}\Bigl[\sup_{u\in[s,s+h]}|X_u^{t,x}-X_s^{t,x}|^p \Bigr]
&\leq& C\Bigl(1+\mbb{E}\Bigl[\sup_{r\in[t,T]}|X_r^{t,x}|^p|\Bigr]\Bigr)h\nn \\
&\leq& C(1+|x|^p)h \nn
\eea
which gives the desired result.\\
(c)
Without loss of generality, we assume $0\leq t^\prime\leq t\leq T$.
We separate the problem into the three cases with respect to the range of $s$.
Firstly, we clearly have
\bea
\mbb{E} \sup_{0\leq s\leq t^\prime} |X_s^{t,x}-X_s^{t^\prime,x^\prime}|^p\leq |x-x^\prime|^p.\nn 
\eea
Secondly, let us consider
\bea
&&\mbb{E}\sup_{t^\prime\leq s\leq t}|X_s^{t,x}-X_s^{t^\prime,x^\prime}|^p
=\mbb{E}\sup_{t^\prime\leq s\leq t}|x-X_s^{t^\prime,x^\prime}|^p\nn \\
&&\leq C \mbb{E}\sup_{t^\prime\leq s\leq t}\Bigl(|x^\prime-X_s^{t^\prime,x^\prime}|^p+|x-x^\prime|^p\Bigr)\nn \\
&&\leq C\bigl(|x-x^\prime|^p+(1+|x^\prime|^p)|t-t^\prime|\bigr)\nn
\eea
where, in the last inequality,  we have used the result (b).

Finally, we consider the  case $s\geq t$. Note that
\bea
X_s^{t^\prime,x^\prime}=X_t^{t^\prime,x^\prime}+\int_t^s b(r,X_r^{t^\prime,x^\prime})dr+\int_t^s \sigma(r,X_r^{t^\prime,
x^\prime})dW_r+\int_t^s\int_E \gamma(r,X_{r-}^{t^\prime,x^\prime},e)\wt{\mu}(dr,de)\nn
\eea
and hence
\bea
X_s^{t,x}-X_s^{t^\prime,x^\prime}&=&x-x^\prime-(X_t^{t^\prime,x^\prime}-x^\prime)\nn \\
&+&\int_t^s [b(r,X_r^{t,x})-b(r,X_r^{t^\prime,x^\prime})]dr+
\int_t^s[\sigma(r,X_r^{t,x})-\sigma(r,X_r^{t^\prime,x^\prime})]dW_r\nn \\
&+&\int_t^s\int_E [\gamma(r,X_{r-}^{t,x},e)-\gamma(r,X_{r-}^{t^\prime,x^\prime},e)]\wt{\mu}(dr,de).\nn
\eea
Applying BDG inequality and Lemma~\ref{ap-lemma-BGJ}, one obtains
\bea
&&\mbb{E}\Bigl[\sup_{s\in[t,T]}|X_s^{t,x}-X_s^{t^\prime,x^\prime}|^p\Bigr]\leq C\mbb{E}\Bigl\{ |x-x^\prime|^p+|X_t^{t^\prime,x^\prime}-x^\prime|^p\nn \\
&&+\Bigl(\int_t^T |b(r,X_r^{t,x})-b(r,X_r^{t^\prime,x^\prime})|dr\Bigr)^p+
\Bigl(\int_t^T |\sigma(r,X_r^{t,x})-\sigma(r,X_r^{t^\prime,x^\prime})|^2dr\Bigr)^{p/2}\nn \\
&&+\int_t^T |X_r^{t,x}-X_r^{t^\prime,x^\prime}|^p dr\Bigr\}\nn \\
&&\leq C(|x-x^\prime|^p+(1+|x^\prime|^p)|t-t^\prime|)+C\int_t^T \mbb{E}\Bigl[ \sup_{s\in[r,T]}|X_s^{t,x}-X_s^{t^\prime,x^\prime}|^p\Bigr] dr \nn
\eea
where, in the last inequality, the result (b) was used.

Using the backward Gronwall inequality, one obtains 
\bea
\mbb{E}\Bigl[\sup_{s\in[t,T]}|X_s^{t,x}-X_s^{t^\prime,x^\prime}|^p\Bigr]\leq C(|x-x^\prime|^p+(1+|x^\prime|^p)|t-t^\prime|). \nn
\eea
Adding the above three cases and flipping the role of $t,t^\prime$, one obtains in general
\bea
\mbb{E}\Bigl[\sup_{s\in[0,T]}|X_s^{t,x}-X_s^{t^\prime,x^\prime}|^p\Bigr]\leq C\Bigl(|x-x^\prime|^p+(1+(|x|\vee |x^\prime|)^p)|t-t^\prime|\Bigr)~.\nn
\eea

%%%%%%%%%%%%%%%%%%%%%%%%%%%%%%%%%%%%%%%%%%%%%%%%%%%%%%%%%%%%%%%%%%
\section{Existence and uniqueness results for Lipschitz case}
%%%%%%%%%%%%%%%%%%%%%%%%%%%%%%%%%%%%%%%%%%%%%%%%%%%%%%%%%%%%%%%%%%
Anticipated BSDEs under the global Lipschitz condition have been 
studied by many authors.
Our setup is a bit different from the standard one, in particular at 
the terminal condition and also at the point where the continuity of the driver 
is defined with respect to the 
uniform norm of the path rather than $\mbb{L}^2[0,T]$-norm.
For readers' convenience, we provide a proof under our particular setup.
It is restricted to the simplest form relevant for our purpose. One can readily 
generalize it to multi-dimensional setups with the future $(Z,\psi)$-dependence (See~\cite{Oksendal} among others.).

Let us consider the ABSDE for $t\in[0,T]$
\bea
Y_t=\xi+\int_t^T \mbb{E}_{\calf_r} f\bigl(r,(Y_v)_{v\in[r,T]},Y_r,Z_r,\psi_r\bigr)dr-\int_t^T Z_r dW_r
-\int_t^T\int_E \psi_r(e)\wt{\mu}(dr,de)
\label{eq-lip-bsde}
\eea
where
$f:\Omega\times [0,T]\times \mbb{D}[0,T]\times \mbb{R}\times \mbb{R}^{1\times d}\times \mbb{L}^2(E,\nu)\rightarrow \mbb{R}$
and  $\xi$ is an $\calf_T$-measurable random variable.

\begin{assumption}
\label{assumption-Lipschitz}
(i) The driver $f$ is a map such that for every $(y,z,\psi)\in\mbb{R}\times \mbb{R}^{1\times d}\times \mbb{L}^2(E,\nu)$
and any \cadlag $\mbb{F}$-adapted process $(Y_v)_{v\in[0,T]}$, the process $\bigl(\mbb{E}_{\calf_t}f(t,(Y_v)_{v\in[t,T]},y,z,\psi), t\in[0,T]\bigr)$ is progressively measurable. \\
(ii) For every $(q,y,z,\psi),(q^\prime,y^\prime,z^\prime,\psi^\prime) \in \mbb{D}[0,T]\times \mbb{R}\times \mbb{R}^{1\times d}
\times \mbb{L}^2(E,\nu)$, there exists some positive constant $K$ such that
\bea
&&\bigl| f\bigl(t,(q_v)_{v\in[t,T]},y,z,\psi\bigr)-f\bigl(t,(q^\prime_v)_{v\in[t,T]},y^\p,z^\p,\psi^\p\bigr)\bigr| \nn \\
&&\quad \leq K\Bigl(\sup_{v\in[t,T]}|q_v-q^\p_v|+|y-y^\p|+|z-z^\p|+||\psi-\psi^\p||_{\mbb{L}^2(\nu)}\Bigr)\nn
\eea
$d\mbb{P}\otimes dt$-a.e. $(\omega,t)\in\Omega\times [0,T]$.\\
(iii) $\mbb{E}\Bigl[|\xi|^2+\Bigl(\int_0^T|f(r,0,0,0,0)|dr\Bigr)^2\Bigr]<\infty$.
\end{assumption}

\begin{proposition}
\label{prop-Lipschitz}
Under Assumption~\ref{assumption-Lipschitz}, there exists a unique solution $(Y,Z,\psi)\in \mbb{S}^2\times \mbb{H}^2\times \mbb{J}^2$
to the ABSDE (\ref{eq-lip-bsde}).
\begin{proof}
We prove the claim by constructing a strictly contracting map $\Phi:\calk^2[0,T]\ni(Y^k,Z^k,\psi^k)\mapsto \Phi(Y^k,Z^k,\psi^k)=:(Y^{k+1},Z^{k+1},\psi^{k+1})\in \calk^2[0,T]$ defined by
\bea
Y^{k+1}_t=\xi+\int_t^T \mbb{E}_{\calf_r}f\bigl(r,(Y^k_v)_{v\in[r,T]},Y^k_r,Z^k_r,\psi^k_r\bigr)dr
-\int_t^T Z_r^{k+1}dW_r-\int_t^T \int_E \psi_r^{k+1}(e)\wt{\mu}(dr,de)\nn
\eea
with $k\in\mbb{N}_0$ and $(Y^0,Z^0,\psi^0)\equiv (0,0,0)$. It is easy to see that the map is well-defined. Let
\be
\del Y^{k+1}:=Y^{k+1}-Y^k, \quad \del Z^{k+1}:=Z^{k+1}-Z^k, \quad \del \psi^{k+1}:=\psi^{k+1}-\psi^k,\quad \Theta^k:=(Y^k,Z^k,\psi^k)~.\nn
\ee
We consider the norm $||\cdot||_{\calk^2_\beta}$ equivalent to $||\cdot||_{\calk^2}$ defined with some $\beta >0$
\bea
||(Y,Z,\psi)||^2_{\calk^2_\beta}:=\mbb{E}\Bigl[\sup_{r\in[0,T]}|e^{\beta r}Y_r|^2\Bigr]+
\mbb{E}\int_0^T |e^{\beta r}Z_r|^2 dr+\mbb{E}\int_0^T ||e^{\beta r}\psi_r||^2_{\mbb{L}^2(\nu)}dr~.\nn
\eea
Applying Ito formula to $e^{2\beta t}|\del Y_t^{k+1}|^2$, one obtains for any $t\in[0,T]$
\bea
&&e^{2\beta t}|\del Y_t^{k+1}|^2+\int_t^T e^{2\beta r}|\del Z_r^{k+1}|^2 dr+\int_t^T \int_E e^{2\beta r}
|\del \psi_r^{k+1}(e)|^2\mu(dr,de)\nn \\
&&=\int_t^T e^{2\beta r}\Bigl(2\del Y_r^{k+1}\mbb{E}_{\calf_r}\bigl[
f(r,(Y^k_v)_{v\in[r,T]},\Theta_r^k)-f(r,(Y^{k-1}_v)_{v\in[r,T]},\Theta_r^{k-1})\bigr]-2\beta |\del Y_r^{k+1}|^2\Bigr)dr\nn\\
&&-\int_t^T e^{2\beta r}2\del Y_r^{k+1}\del Z_r^{k+1}dW_r-\int_t^T \int_E e^{2\beta r}
2\del Y_{r-}^{k+1}\del \psi_r^{k+1}(e)\wt{\mu}(dr,de)~.
\label{eq-beta-ysq}
\eea
For any $\ep>0$, one has
\bea
&&2\del Y_r^{k+1}\mbb{E}_{\calf_r}\bigl[
f(r,(Y^k_v)_{v\in[r,T]},\Theta_r^k)-f(r,(Y^{k-1}_v)_{v\in[r,T]},\Theta_r^{k-1})\bigr]-2\beta |\del Y_r^{k+1}|^2\nn \\
&&\leq 2K|\del Y_r^{k+1}|\Bigl(2\mbb{E}_{\calf_r}\bigl[||\del Y^k||_{[r,T]}\bigr]+|\del Z_r^k|+||\del \psi_r^k||_{\mbb{L}^2(\nu)}\Bigr)-2\beta 
|\del Y_r^{k+1}|^2\nn\\
&&\leq \Bigl(\frac{6K^2}{\ep}-2\beta\Bigr)|\del Y_r^{k+1}|^2+\ep\Bigl(
\mbb{E}_{\calf_r}\bigl[||\del Y^k||_{[r,T]}^2\bigr]+|\del Z_r^k|^2+||\del \psi_r^{k}||^2_{\mbb{L}^2(\nu)}\Bigr)~.\nn
\eea
Thus, choosing $\beta=\beta(\ep)=3K^2/\ep$ and taking expectation with $t=0$ yields
\bea
||e^{\beta \cdot}\del Z^{k+1}||^2_{\mbb{H}^2}+||e^{\beta\cdot}\del \psi^{k+1}||^2_{\mbb{J}^2}
\leq \ep\Bigl(T||e^{\beta\cdot}\del Y^k||^2_{\mbb{S}^2}+||e^{\beta\cdot \del} \del Z^k||^2_{\mbb{H}^2}+
||e^{\beta\cdot}\del \psi^{k}||^2_{\mbb{J}^2} \Bigr)~.
\label{eq-z-psi-estimate-L}
\eea 

Next,  let us apply the BDG inequality (Theorem 48 in IV.4. of \cite{Protter}) to (\ref{eq-beta-ysq}).
Then there exists some constant $C$ such that
\bea
&&\mbb{E}\Bigl[||e^{\beta \cdot} \del Y^{k+1}||^2_{[0,T]}\Bigr]
\leq \ep \Bigl(T||e^{\beta\cdot}\del Y^k||^2_{\mbb{S}^2}+||e^{\beta\cdot \del} \del Z^k||^2_{\mbb{H}^2}+
||e^{\beta\cdot}\del \psi^{k}||^2_{\mbb{J}^2} \Bigr)\nn \\
&&+C\mbb{E}\Bigl[\Bigl(\int_0^T |e^{\beta r}\del Y_r^{k+1}|^2|e^{\beta r}\del Z_r^{k+1}|^2dr\Bigr)^\frac{1}{2}\Bigr]+
C\mbb{E}\Bigl[\Bigl(\int_0^T\int_E |e^{\beta r}\del Y_{r-}^{k+1}|^2|e^{\beta r}\del \psi_r^{k+1}(e)|^2\mu(dr,de)\Bigr)^\frac{1}{2}
\Bigr]\nn \\
&&\leq \ep \Bigl(T||e^{\beta\cdot}\del Y^k||^2_{\mbb{S}^2}+||e^{\beta\cdot \del} \del Z^k||^2_{\mbb{H}^2}+
||e^{\beta\cdot}\del \psi^{k}||^2_{\mbb{J}^2} \Bigr)+\frac{1}{2}\mbb{E}\Bigl[||e^{\beta\cdot}\del Y^{k+1}||^2_{[0,T]}\Bigr] \nn \\
&&+C\Bigl(||e^{\beta \cdot}\del Z^{k+1}||^2_{\mbb{H}^2}+||e^{\beta\cdot}\del\psi^{k+1}||^2_{\mbb{J}^2}\Bigr)\nn~.
\eea
Thus, with some constant $C$ (which is independent of $\ep,\beta$), 
\be
||e^{\beta\cdot} \del Y^{k+1}||^2_{\mbb{S}^2}\leq 2\ep\Bigl(T||e^{\beta\cdot}\del Y^k||^2_{\mbb{S}^2}+||e^{\beta\cdot \del} \del Z^k||^2_{\mbb{H}^2}+
||e^{\beta\cdot}\del \psi^{k}||^2_{\mbb{J}^2} \Bigr)+C\Bigl(||e^{\beta \cdot}\del Z^{k+1}||^2_{\mbb{H}^2}+||e^{\beta\cdot}\del\psi^{k+1}||^2_{\mbb{J}^2}\Bigr)~. \nn
\ee
Combining with (\ref{eq-z-psi-estimate-L}), one obtains
\bea
\bigl|\bigl|(\del Y^{k+1},\del Z^{k+1},\del\psi^{k+1})\bigr|\bigr|^2_{\calk^2_{\beta(\ep)}}
\leq \ep(C+3)(T\vee 1)\bigl|\bigl|(\del Y^{k},\del Z^{k},\del\psi^{k})\bigr|\bigr|^2_{\calk^2_{\beta(\ep)}}\nn
\eea
and hence by choosing $\ep$ so that $\ep(C+3)(T\vee 1)<1$ (and $\beta(\ep)$ accordingly) makes the map $\Phi$ strict contraction with respect to the norm $\calk^2_{\beta(\ep)}$.
This proves the existence as well as the uniqueness.
\end{proof}
\end{proposition}

%%%%%%%%%%%%%%%%%%%%%%%%%%%%%%%%%%%%%%%%%%%%%%%%%%%%%%%%%%%%%%%%%%%%%%%%%%%%%%%%%
\section{Comparison principle for non-anticipated settings}
%%%%%%%%%%%%%%%%%%%%%%%%%%%%%%%%%%%%%%%%%%%%%%%%%%%%%%%%%%%%%%%%%%%%%%%%%%%%%%%%%
Consider the two BSDEs with $i=\{1,2\}$,
\bea
Y_t^i=\xi_i+\int_t^T \wt{f}_i(r,Y^i_r,Z_r^i,\psi^i_r)dr-\int_t^T Z_r^i dW_r-\int_t^T\int_E \psi_r^i(e)\wt{\mu}(dr,de)
\label{eq-nonA}
\eea
for $t\in[0,T]$.
%%%%%%%%%%%%%%%%%%%%%%%%%%%%%%%%
\begin{lemma}
\label{lemma-comparison-nonA}
Suppose $(\xi,\wt{f}_i)_{1\leq i\leq 2}$ satisfy Assumptions 3.1, 3.2 and 4.1 of \cite{FT-Qexp}, which correspond
to Assumptions~\ref{assumption-structure-nonM}, \ref{assumption-LLC-nonM} and \ref{assumption-A-gamma} of the current paper without the 
$Y$'s future path dependence, respectively. If $\xi_1 \leq \xi_2$ a.s. and $\wt{f}_1(r,y,z,\psi)\leq \wt{f}_2(r,y,z,\psi)$ $d\mbb{P}\otimes dr$-a.e. for every
$(y,z,\psi)\in \mbb{R}\times \mbb{R}^{1\times d}\times \mbb{L}^2(E,\nu)$, 
then $Y_t^1\leq Y_t^2$ $\forall t\in[0,T]$ a.s.
\begin{proof}
One can prove it in the same way as Theorem 2.5 of \cite{Royer}. 
By Theorem 4.1~\cite{FT-Qexp}, there exists a unique solution $(Y^i,Z^i,\psi^i)_{1\leq i\leq 2}\in \mbb{S}^\infty\times 
\mbb{H}^2_{BMO}\times \mbb{J}^2_{BMO}$ to the BSDEs (\ref{eq-nonA}) satisfying the universal bounds.
Let us put $\del Y:=Y^1-Y^2$, $\del Z:=Z^1-Z^2$, $\del\psi:=\psi^1-\psi^2$,
$\del \wt{f}(r):=(\wt{f}_1-\wt{f}_2)(r,Y_r^1,Z_r^1,\psi_r^1)$.
We also introduce the two progressively measurable processes $(a_r)_{r\in[0,T]}$, $(b_r)_{r\in[0,T]}$
given by
\bea
\hspace{-5mm}a_r:=\frac{\wt{f}_2(r,Y_r^1,Z_r^1,\psi_r^1)-\wt{f}_2(r,Y^2_r,Z_r^1,\psi_r^1)}{\del Y_r}\bold{1}_{\del Y_r\neq 0},~
b_r:=\frac{\wt{f}_2(r,Y_r^2,Z_r^1,\psi_r^1)-\wt{f}_2(r,Y^2_r,Z_r^2,\psi_r^1)}{|\del Z_r|^2}
\bold{1}_{\del Z_r\neq 0}\del Z_r^\top\nn.
\eea
Note that $a\in \mbb{S}^{\infty}$ and $b\in \mbb{H}^2_{BMO}$ due to the universal bounds and the local Lipschitz continuity.
By Assumption 4.1 of \cite{FT-Qexp}, which is the $A_\Gamma$-condition, there exists a 
$\mbb{P}\otimes \cale$-measurable process $\Gamma$ such that
\bea
\del Y_t&\leq& \del \xi+\int_t^T \Bigl(\del \wt{f}(r)+a_r \del Y_r+ b_rZ_r+\int_\mbb{E}\Gamma_r(e)\del\psi_r(e)\nu(de)\Bigr)dr\nn \\
&&-\int_t^T \del Z_r dW_r-\int_t^T \int_E \del \psi_r(e)\wt{\mu}(dr,de)
\label{eq-comp-2}
\eea
satisfying $C_1(1\wedge |e|)\leq |\Gamma(e)|\leq C_2(1\wedge |e|)$ with some constant $C_1>-1$ and $C_2\geq 0$.
Here the fact that $Y^i\in\mbb{S}^\infty, \psi^i\in \mbb{J}^{\infty}$ was used.
Since $\calm:=\int_0^\cdot b_r^\top dW_r+\int_0^\cdot\int_E \Gamma_r(e)\wt{\mu}(dr,de)$ is a BMO-martingale with jump size strictly bigger than $-1$, one can define an equivalent measure $\mbb{Q}$ by $d\mbb{Q}/d\mbb{P}=\cale_T(\calm)$.
Thus one obtains from (\ref{eq-comp-2}) 
\bea
\del Y_t\leq \mbb{E}^{\mbb{Q}}_{\calf_t}\Bigl[e^{R_{t,T}}\del \xi+\int_t^T e^{R_{t,r}}\del\wt{f}(r)dr\Bigr]\nn
\eea
with $R_{t,s}:=\int_t^s a_r dr$. This proves the claim.
\end{proof}
\end{lemma}

\end{appendix}

%%%%%%%%%%%%%%%%%%%%%%%%%%%%%%%%%%%%%%%%%
\section*{Acknowledgement}
The research is partially supported by Center for Advanced Research in Finance (CARF).

%%%%%%%%%%%%%%%%%%%%%%%%%%%%%%%%%%%%%%%%%%%%%%%%%%%%%%%%%%%%%%%%%%%%%

%%%%%%%%%%%%%%%%%%%%%%%%%%%%%%%%%%%%%%%%%%%%%%%%%%%%%%%%%%%%%%%%%%%%

\end{document}

%% file: Fmacro-2015.tex
%nakamacro.tex(H120522;0730)
%\documentstyle[11pt]{article}
%\setlength{\textwidth}{10.5in}
%\setlength{\oddsidemargin}{0in}
%\setlength{\topmargin}{-0.52in}
%\setlength{\textheight}{9.0in}
%\setlength{\footskip}{0.7in}

\newtheorem{definition}{Definition}[section]
\newtheorem{assumption}{Assumption}[section]
\newtheorem{condition}{$[$ C}
\newtheorem{lemma}{Lemma}[section]
\newtheorem{proposition}{Proposition}[section]
\newtheorem{theorem}{Theorem}[section]
\newtheorem{remark}{Remark}[section]
\newtheorem{example}{Example}[section]
\newtheorem{corollary}{Corollary}[section]
%--------------------------------------------------------------------------
%BOLD FACES
\def\n{{\bf n}}
\def\A{{\bf A}}
\def\B{{\bf B}}
\def\C{{\bf C}}
\def\D{{\bf D}}
\def\E{{\bf E}}
\def\F{{\bf F}}
\def\G{{\bf G}}
\def\H{{\bf H}}
\def\I{{\bf I}}
\def\J{{\bf J}}
\def\K{{\bf K}}
\def\L{{\bf L}}
\def\M{{\bf M}}
\def\N{{\bf N}}
\def\O{{\bf O}}
\def\P{{\bf P}}
\def\Q{{\bf Q}}
\def\R{{\bf R}}
\def\S{{\bf S}}
\def\T{{\bf T}}
\def\U{{\bf U}}
\def\V{{\bf V}}
\def\W{{\bf W}}
\def\X{{\bf X}}
\def\Y{{\bf Y}}
\def\Z{{\bf Z}}
\def\cala{{\cal A}}
\def\calb{{\cal B}}
\def\calc{{\cal C}}
\def\cald{{\cal D}}
\def\cale{{\cal E}}
\def\calf{{\cal F}}
\def\calg{{\cal G}}
\def\calh{{\cal H}}
\def\cali{{\cal I}}
\def\calj{{\cal J}}
\def\calk{{\cal K}}
\def\call{{\cal L}}
\def\calm{{\cal M}}
\def\caln{{\cal N}}
\def\calo{{\cal O}}
\def\calp{{\cal P}}
\def\calq{{\cal Q}}
\def\calr{{\cal R}}
\def\cals{{\cal S}}
\def\calt{{\cal T}}
\def\calu{{\cal U}}
\def\calv{{\cal V}}
\def\calw{{\cal W}}
\def\calx{{\cal X}}
\def\caly{{\cal Y}}
\def\calz{{\cal Z}}
%
%YOKUTUKAUMONO
\def\sskip{\hspace{0.5cm}}
\def\simleq{ \raisebox{-.7ex}{\em $\stackrel{{\textstyle <}}{\sim}$} }
\def\leqsim{ \raisebox{-.7ex}{\em $\stackrel{{\textstyle <}}{\sim}$} }
\def\ep{\epsilon}
\def\half{\frac{1}{2}}
\def\iku{\rightarrow}
\def\Iku{\Rightarrow}
\def\ikup{\rightarrow^{p}}
\def\inclusion{\hookrightarrow}
\def\cadlag{c\`adl\`ag\ }
\def\up{\uparrow}
\def\down{\downarrow}
\def\doti{\Leftrightarrow}
\def\douti{\Leftrightarrow}
\def\dochi{\Leftrightarrow}
\def\douchi{\Leftrightarrow}%
%KAIGYOU,ARRAY
\def\yy{\\ && \nonumber \\}
\def\y{\vspace*{3mm}\\}
\def\nn{\nonumber}
\def\be{\begin{equation}}
\def\ee{\end{equation}}
\def\bea{\begin{eqnarray}}
\def\eea{\end{eqnarray}}
\def\beas{\begin{eqnarray*}}
\def\eeas{\end{eqnarray*}}
%
%KONO RONBUN DE TUKAU MONO
\def\hd{\hat{D}}
\def\hv{\hat{V}}
\def\hsd{{\hat{d}}}
\def\hx{\hat{X}}
\def\hsx{\hat{x}}
\def\bsx{\bar{x}}
\def\bsd{{\bar{d}}}
\def\bx{\bar{X}}
\def\ba{\bar{A}}
\def\bb{\bar{B}}
\def\bc{\bar{C}}
\def\bv{\bar{V}}
\def\balpha{\bar{\alpha}}
\def\bbalpha{\bar{\bar{\alpha}}}
\def\combi{\l(\begin{array}{c}\alpha\\ \beta \end{array}\r)}
\def\f{^{(1)}}
\def\s{^{(2)}}
\def\ss{^{(2)*}}
\def\l{\left}
\def\r{\right}
\def\a{\alpha}
\def\b{\beta}
\def\L{\Lambda}
%上に定義されたコマンドは数式モ−ドで用いる。
%--------------------------------------------------